\documentstyle[preprint,aps,eqsecnum,epsfig]{revtex}
\tighten
\begin{document}
\preprint{PITT-98-; CMU-HEP-98-; LPTHE-98/45} 
\draft 
\title{\bf NON-EQUILIBRIUM BOSE-EINSTEIN 
CONDENSATES, DYNAMICAL SCALING AND SYMMETRIC EVOLUTION IN THE LARGE N
$ \Phi^4 $ THEORY}
\author{{\bf D. Boyanovsky$^{(a)}$, H. J. de Vega$^{(b)}$,
R. Holman$^{(c)}$ and 
J. Salgado$^{(b)}$ }} \address { (a)Department of Physics and Astronomy,
University of Pittsburgh, Pittsburgh, PA 15260 USA\\ (b)
LPTHE\footnote{Laboratoire Associ\'{e} au CNRS UMR 7589.}  Universit\'e Pierre
et Marie Curie (Paris VI) et Denis Diderot (Paris VII), Tour 16, 1er. \'etage,
4, Place Jussieu 75252 Paris, Cedex 05, France \\ 
(c) Department of Physics,
Carnegie Mellon University, Pittsburgh, PA 15213, USA} \date{\today} \maketitle
\begin{abstract}

We analyze the non-equilibrium dynamics of the $ O(N) \; \Phi^4 $ model
in the large N limit with broken symmetry tree level potential and for
states of large energy density. The dynamics is dramatically different
when the energy density is above the top of the tree level 
potential $V_0$ than when it is below it. When the energy density
is below $ V_0 $, we find that non-perturbative particle production through 
spinodal instabilities provides a dynamical mechanism for the Maxwell
construction. The 
asymptotic values of the order parameter only depend on the initial
energy density and all values between the minima of the tree level
potential are available, the asymptotic {\em dynamical}
`effective potential' is flat
between the minima. When the energy density is larger than $ V_0 $, the
evolution samples ergodically the broken symmetry states, as a consequence of 
non-perturbative particle production via parametric amplification. 
 Furthermore, we examine the  quantum dynamics of phase ordering into
the broken symmetry phase and 
find novel scaling behavior of the correlation function.
There is a  crossover in the dynamical correlation length 
at a time scale $ t_s \approx \ln(1/\lambda) $. For $ t<t_s $ the dynamical
correlation length $ \xi(t) \propto \sqrt{t} $ and the evolution is
dominated by linear instabilities and spinodal decomposition, whereas
for $ t>t_s $  the evolution is non-linear and dominated by the onset of
 non-equilibrium   Bose-Einstein condensation of long-wavelength
Goldstone bosons. In this regime a true scaling solution emerges with a
non-perturbative anomalous scaling length dimension $z=1/2$ and a
  dynamical correlation length $\xi(t) \propto (t-t_s)$. The equal time correlation function in this scaling regime 
vanishes for $r>2(t-t_s)$ by causality. For $t>t_s$ phase ordering
proceeds by the formation of domains that grow at the speed of light,
with  non-perturbative condensates of Goldstone
bosons and the equal time correlation function falls of as $ 1/r $. A
semiclassical but stochastic description emerges for time scales $ t > t_s$.
Our results are compared to phase ordering in {\em classical} stochastic
descriptions in condensed matter and cosmology. 
 
\end{abstract}

\section{Introduction}

Recent studies of non-equilibrium evolution of high energy density states have
revealed novel and unexpected phenomena ranging from strong dissipative
processes via particle production\cite{dis,linon} to novel aspects of symmetry
breaking\cite{mottola,tsu}. These non-equilibrium effects are crucial to
understanding the end of inflation and the reheating
problem\cite{linon,lindekov,big}. They are also conjectured to be relevant for
baryogenesis, the partial restoration of symmetries during phase
transitions\cite{kolb,tkachev}, for the formation of defects during
non-equilibrium phase transitions\cite{tkachev,kawa} and reheating dynamics in
gauge and fermionic theories\cite{baacke1}.  More recently a novel form of
anomalous relaxation with non-universal critical exponents\cite{late} has been
found in scalar field theories due to non-equilibrium quantum effects.

One of the important aspects of these phenomena are their non-perturbative
character. There are few consistent approaches that can be used for
their description. While a classical approach has been advocated to address
these issues\cite{lindekov,kolb,tkachev,kawa} we have emphasized the
application of the large $ N $ limit that captures both the quantum aspects as
well as the emergence of classicality\cite{dis,linon,mottola,tsu,big,late}.
The large $ N $ limit in scalar field theories has a consistent implementation
within the non-equilibrium quantum field theory
formulation\cite{dis,linon,losa} and the renormalization aspects had been
studied in detail\cite{baacke2,FRW}.  Furthermore it provides a consistent,
renormalizable framework to study non-perturbative phenomena that can be
systematically improved\cite{dis,linon,losa}.

In this article we focus on further new non-equilibrium
aspects. First, we discuss how non-perturbative particle production due to
spinodal instabilities gives rise to a {\em dynamical} version of the Maxwell
construction of statistical mechanics. It is a well known result that the
static effective potential becomes complex in the region of the phase diagram in which
coexistence of phases can occur. The imaginary part has been deemed to arise
from a failure of the loop expansion, but a more careful treatment reveals that
this imaginary part is actually related to the decay rate of a quantum
mechanically unstable state\cite{erickwu,boyvega}. More precisely, the
static effective potential approach is just not appropriate to
study the dynamic of phase transitions. At the level of formal
statistical mechanics, the mixed phase is described via a Maxwell construction
that leads to an effective potential (free energy) that is flat in the
coexistence region. The Maxwell constructed effective potential has been
confirmed by a careful analysis using the `exact' renormalization
group\cite{rg,polo,tetradis} of the {\em equilibrium} coarse grained free
energy. 

Here we study the dynamical aspects of the Maxwell construction in the large $
N $ limit of a scalar field theory. We begin by reviewing the relevant features
of the static effective potential in the large $ N $ limit\cite{root,aks,barmo}
to highlight the fact that the imaginary part of the effective potential is
{\em not} an artifact of the loop expansion but a more fundamental shortcoming
of a static and equilibrium description of the coexistence region in terms of
an homogeneous order parameter. We then study in detail how non-equilibrium
dynamics, and in particular, particle production due to spinodal instabilities,
leads to a flat dynamic potential in the region of coexistence. We find that
{\em all} expectation values of the scalar field between the minima of the tree
level potential are available asymptotically and that the asymptotic value is
reached with a vanishing  effective mass for the expectation
value. This is the dynamical equivalent of a `flat {\em dynamical} potential' and the
availability of all possible asymptotic expectation values between the minima
is the dynamical equivalent of the lever rule for coexistence of phases in the
Maxwell construction. For a given 
initial expectation value, such that the total energy density, 
$ \varepsilon $,  is smaller than the maximum of the tree level  
potential $ V(0) $, [here $V(\Phi)$
is the potential for the $ O(N)$ field $ \Phi $], 
we find that the asymptotic expectation value depends on the initial
conditions only through  the total energy as follows
$$
<\Phi({\vec x}, \infty )> = \sqrt{2 \over \lambda} |m| \left[ 1 - {
\varepsilon \over V(0)}\right]^x
$$
where $ \lambda $ and $ m $ are the renormalized quartic self-coupling and
mass respectively and $x$ a new dynamical anomalous exponent which is universal in the weak coupling limit  and given by $ x=0.25 $.

In summary, the {\em dynamical} effective potential is always {\bf real} and is flat
for $ \varepsilon < V(0) $. Two physically different situations appear here: 

a) when the energy density $\varepsilon$ is smaller than the maximum of the
tree level potential, $\varepsilon < V(0)$ (in terms of renormalized
mass and coupling), the non-equilibrium 
dynamical evolution leads to broken symmetry states and the Maxwell
construction  via non-perturbative particle production as a consequence of
the spinodal instabilities.

b) when $\varepsilon > V(0)$, in which case the energy density is larger 
than the maximum of the tree level potential, we find that the evolution
is {\em symmetric} in that the time evolution of the expectation value
of the scalar field samples the minima of the tree level potential
symmetrically and ergodically with vanishing average over time scales
much longer than the period.   
What we find is that this evolution is a consequence of
the growth of the root mean square fluctuations due to parametric amplification
which eventually induces a positive mass squared for the field. We show
numerically that the expectation value relaxes to zero asymptotically
transferring all of the energy to quantum fluctuations via parametric
amplification in accord with the results in\cite{big,late}. 

Our final topic concerns the quantum non-equilibrium dynamics of phase ordering
in the $O(N)$ model. This is an important topic and is the basis of the Kibble
mechanism for defect formation as the early universe cools through phase
transitions\cite{kibble,kibble2,vilen}. Recently, there have been some attempts
at constructing a non-equilibrium, real-time, description of the dynamics of
this process\cite{zurek}. The early stages of the dynamics of phase ordering in
quantum field theory models begin with the growth of spinodally unstable
long-wavelength field modes and a dynamical correlation length
emerges\cite{boylee}. The dynamics of non-equilibrium fluctuations during the
early spinodally unstable stage has been studied and clarified recently
providing a deeper understanding of the quantum fluctuations that trigger the
process of phase ordering\cite{rivers,beilok}. We go beyond the early
and intermediate time regime which is dominated by the linear instabilities,
into the highly non-linear regime which is dominated by non-linear
resonances\cite{late} and that begins at the spinodal time $ t_s \sim m^{-1}
\ln(1/\lambda) $. This time scale determines the emergence of a semiclassical but {\em
stochastic} description which we discuss and clarify. After $ t_s $,
semiclassical, large amplitude ($\propto 1/\sqrt{\lambda}$) field
configurations are represented in the quantum density matrix with unsupressed probability.

We find that for $t> t_s$ a true scaling solution emerges for which
the field acquires a non-perturbative anomalous scaling (length) dimension
$z=1/2$, a dynamical correlation length $\xi(t) \approx (t-t_s)$ emerges. In and the equal-time correlation function vanishes by causality at distances
$ r>2(t-t_s) $. In the non-linear regime we find the onset of a novel form of Bose Einstein
condensation in which the zero momentum mode of the quantum fluctuations grows
asymptotically linearly in time becoming macroscopically populated at long
times. This condensation mechanism results in the equal time correlation
function decreasing as $ 1/r $ for $ r < 2 (t-t_s) $ as a consequence of
massless excitations.
We find a
crossover in the time dependence of the dynamical correlation length at the
spinodal time $ t_s $:
$$
\xi(t) \approx \sqrt{t}\quad  \mbox{for} \quad
t< t_s \quad ; \quad \xi(t)\approx t -t_s\; \quad \mbox{for} \quad t>t_s \; .
$$
The process of phase ordering in this regime is described by the formation of
domains that grow at the speed of light, inside of which there is a
non-perturbative condensate of Goldstone bosons.

In section II we review the construction of the effective potential in the
large $ N $ limit and highlight several important features that are relevant
for the dynamics namely

\begin{itemize}
\item {the regime of validity of the scalar field theory as an
effective theory below the energy scale of the Landau pole $ \approx m \;
e^{16\pi^2/\lambda}\; $, and the consistency of studying this theory for energy
densities $ \propto m^4/\lambda $,} 

\item{ the complex result for the effective
potential in the coexistence region and the failure of such effective potential
to describe dynamical aspects.  In the large $ N $ limit, this effective
potential is {\em exact} and this complex value cannot be blamed on a failure
of the loop expansion but is a consequence of more fundamental shortcomings of
an effective potential description.}
\end{itemize}

In section III we set up our study of the non-equilibrium dynamics in terms of
the quantum density matrix in the large $ N $ limit and its probabilistic
interpretation. This formulation is the most enlightening to use use in order
to discuss the emergence of classicality in terms of probability
distributions. It also provides a natural framework in which to interpret the
formation of semiclassical, large amplitude, field configurations that are
represented in the quantum ensemble with unsuppressed probabilities.

Section IV is devoted to a detailed study of the dynamics in a
broken symmetry potential. Here we distinguish between symmetric and
non-symmetric evolution and provide a simple criterion for ergodic symmetric
evolution in a broken symmetry potential. We contrast the energy
with the effective potential and show that particle production provides a dynamical
mechanism for the Maxwell construction. Section VI deals with the 
non-equilibrium process of phase ordering both in the linear and in the
non-linear regime. We then compare our results to those of phase
ordering kinetics in condensed matter physics\cite{bray,marco} and the scaling
solution found in radiation and matter dominated Friedmann-Robertson-Walker
cosmologies for the non-linear sigma model\cite{turok,filipe}.
We summarize our results and conclusions in section VII.

\section{The Effective Potential in the large $N$ limit}

One of the main conclusions of our work is that the understanding of field
theories to be found in the effective potential can be woefully lacking under
certain circumstances. In order to better compare and contrast our dynamical
results with what would have been expected from an effective potential
analysis, and since there are some subtleties involved in the discussion of the
large $N$ limit effective potential\cite{root,aks,barmo}, we recapitulate this
discussion below.

There are two facets of the $O(N)$ effective potential that we  want to make
contact with: 

\begin{itemize}

\item {Scalar field theories can only be taken as effective field theories with
a restricted regime of validity. The $O(N)$ model is known to have a Landau
ghost at an energy scale $\approx m \; e^{16\pi^2 / \lambda} $, with $m \; ; \;
\lambda$ the renormalized mass and coupling respectively, and can only be used
reliably below this energy scale. A fuller analysis of the $O(N)$ effective
potential in the large $N$ limit determines the range of energy densities for
which the theory can make sensible predictions.}

\item{ The effective potential is {\em complex} in the spinodal region. This is
what gives rise to the standard Maxwell construction. We  want to
understand the {\em dynamics} associated with this construction.}  
\end{itemize}

We consider the  $ O(N) $ vector model with Lagrangian density  given by,

\begin{eqnarray}
{\cal{L}} & =  & \frac{1}{2}\,\dot{\vec{\Phi}}(\vec{x},t)^2-\frac{1}{2}\,
(\vec{\nabla}\vec{\Phi}(\vec{x},t))^2-V(\vec{\Phi}(\vec{x},t))
 \nonumber \\
V(\vec{\Phi})   & =  & \frac{1}{2}m^2_0 \, \vec{\Phi}^2+
\frac{\lambda_0}{8N}\,(\vec{\Phi}^2)^2 +{N\, m^4_0\over 2 \, \lambda_0}
\label{barepotential} 
\end{eqnarray}

To analyze the large $ N $ limit it proves convenient to introduce a
Lagrange multiplier field $ \alpha(x) $\cite{losa} and write 
the above Lagrangian density in the form

$$
{\cal{L}}  =   \frac{1}{2}\,\dot{\vec{\Phi}}(\vec{x},t)^2-\frac{1}{2}\,
[\vec{\nabla}\vec{\Phi}(\vec{x},t)]^2+ \frac{N\;}{2\lambda_0}\;
(\alpha(\vec{x},t)-2\,m^2_0)\;\alpha(\vec{x},t)-
\frac{1}{2}\,\vec{\Phi}(\vec{x},t)^2\;\alpha(\vec{x},t)
$$

The effective potential ($ V_{eff} $) can be obtained in the $ N = \infty $
limit in the following manner\cite{root,aks,barmo}: introducing the expectation
value of the scalar field as $ \phi(\vec{x},t)^2 \equiv \frac1{N} <
\Phi(\vec{x},t)^2 > $, the quadratic functional integral over the field $
\Phi(\vec{x},t) $ can be performed exactly leading to an effective functional
of the Lagrange multiplier $ \alpha(\vec{x},t) $ and the expectation value $
\phi(\vec{x},t)^2 $. The leading order in the large $ N $ limit is obtained via
the saddle point approximation in the functional variable $ \alpha(\vec{x},t) $
and expansion around the saddle point generates a series in powers of $ 1/N
$. Translational invariance dictates that the relevant saddle point solution
must be a space-time constant.  In the large $ N $ limit only coupling and mass
renormalization are needed:
\begin{eqnarray}
\frac{m^2_0}{\lambda_0} & = &  \frac{m^2}{\lambda} -
\frac{\Lambda^2}{16\pi^2} \label{massrenor} \\ 
\frac{1}{\lambda_0} & = & \frac{1}{\lambda}-\frac{1}{32\pi^2}
\ln\left[\left(\frac{2\Lambda}{|m|}\right)^2\frac{1}{e}\right],
\label{coupren}
\end{eqnarray}
where $\Lambda$ is an ultraviolet cutoff and the finite part in the
renormalization of the coupling (\ref{coupren}) has been conveniently chosen to
simplify the expressions below.  As a function of $\alpha(x) = M^2$, and
$\phi^2$, the effective potential at leading order in the large $ N $ limit
is:\cite{root,aks,barmo}
\begin{equation}
V_{eff}( \phi^2 ,M^2) = \frac{M^2}{2} \left[ \phi^2 -{{M^2-
2m^2}\over{\lambda}}  - {{M^2} \over{32\pi^2}}
\log\left({{\sqrt{e}\,|m|^2}\over{M^2}}\right)
\right], \label{leadordveff} 
\end{equation}
where $m$ and $\lambda$ are the renormalized mass and coupling, and we
have chosen $|m|^2$ to be the  renormalization scale. 
Extremizing with respect to the saddle point solution for the Lagrange
multiplier, $M^2$ we are led to the gap equation\cite{root,aks,barmo}:
\begin{equation}\label{gap}
\phi^2(M^2) = {{2(M^2 - m^2)}\over{\lambda}} + 
{{M^2}\over{16\pi^2}}\log\left({{|m|^2}\over{M^2}}\right)
\end{equation}

Shifting the Lagrange multiplier field $\alpha(x)$ by the saddle 
point value $M^2$ we identify $M^2$ as the effective squared mass of
the fluctuations of the  $ \vec{\Phi} $ field. 

We see from eq.(\ref{gap}) that $ \phi^2(M^2) $ increases with $ M^2 $
for $ M^2 < |m|^2 \; e^{ 32\pi^2/\lambda - 1 } $. 

The effective potential {\em at} the solution of the gap equation becomes
\begin{equation}\label{potef}
V_{eff}( \phi^2(M^2) ,M^2) = \frac{M^4}{16} \left[ \frac8{\lambda} + 
{1 \over{4\pi^2}} \log\left({{|m|^2}\over{\sqrt{e}\,M^2}}\right)\right]
\end{equation}

\begin{figure}
\centerline{ \epsfig{file=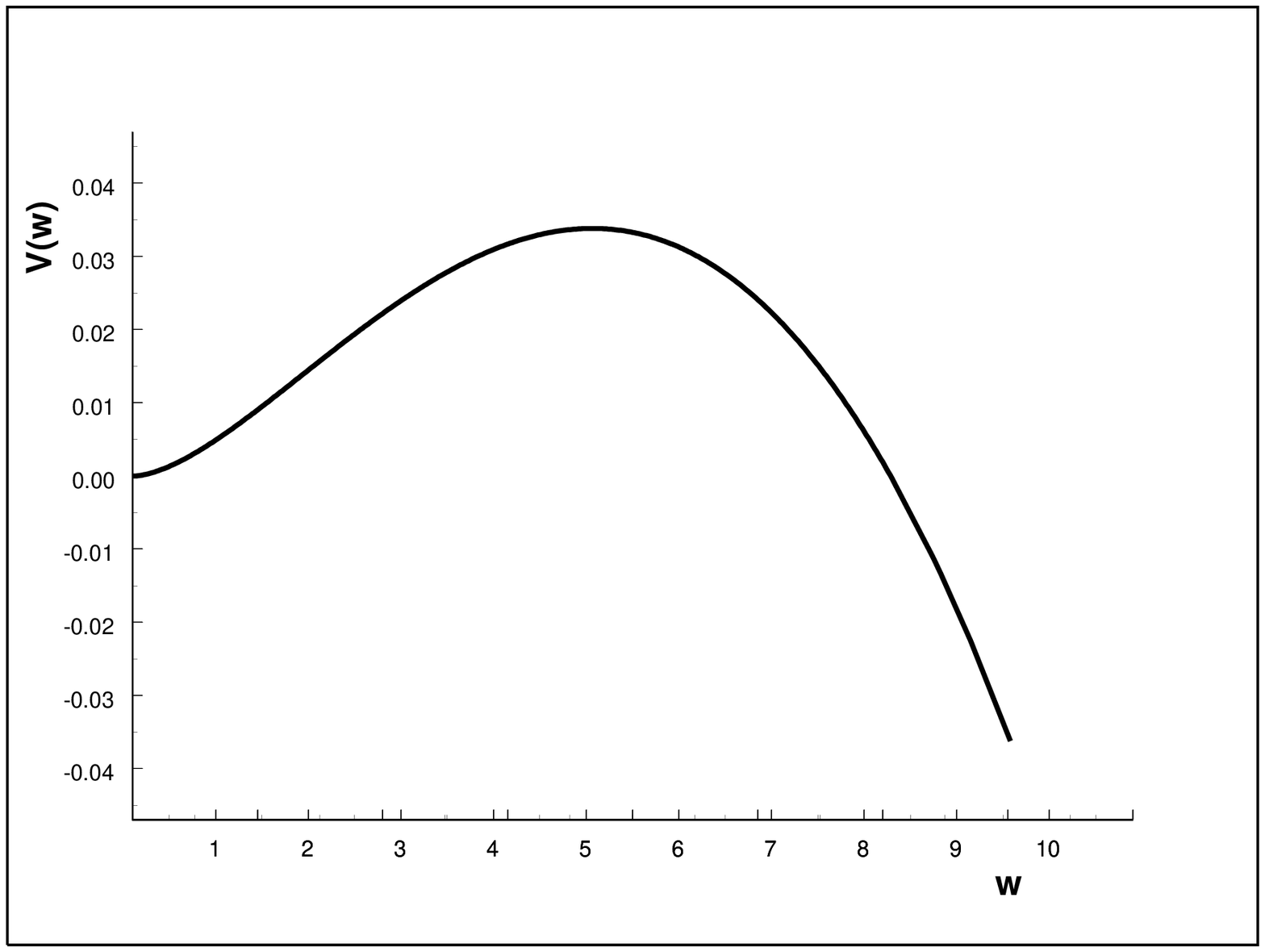,width=6in,height=6in}}
\caption{The dimensionless effective potential $ v(w) $ as a function of the
dimensionless mass $ w $. See eq.(\ref{potv}). 
\label{fig1}}
\end{figure}

It proves convenient to introduce the following dimensionless variables:
\begin{eqnarray}
w & \equiv &   e^{-32\pi^2/\lambda} \; {M^2 \over |m|^2} \nonumber \\
v(w) & \equiv &  {32\pi^2 \over m^4}\; e^{-64\pi^2/\lambda} \; V_{eff}(
\phi^2(M^2) ,M^2) \label{dimlessveff} \\
 \varphi^2 & \equiv & {16 \pi^2 \over |m|^2} \;
e^{-\frac{32\pi^2}{\lambda}}\; \phi^2 \label{finue} 
\end{eqnarray}
in terms of which 
\begin{equation}\label{potv}
v(w) = -\frac12 w^2 \left( \frac12 + \log w \right) \; .
\end{equation}
In fig. \ref{fig1} we plot the dimensionless function $v(w)$
vs. $w$. Furthermore at the solution of the gap equation we find the relation
\begin{equation}\label{fiw}
\varphi^2(w) = -  {32 \pi^2  \over \lambda}\mbox{sign}(m^2)  \;
e^{-32\pi^2/\lambda} - w \log w \; .
\end{equation}
In the regime of interest where $ \lambda/(32 \pi^2) $ is small, the
dimensionless variables $ v, w $ and $ \varphi $ contain the very small factor
$ \, e^{-32\pi^2/\lambda} \, $ times factors that are typically of order
one. Hence, the physically interesting domain of values for $ v, w $ and $
\varphi $ is near the origin. Only for energies near the Landau ghost, $ v, w $
and $ \varphi $ become of order one.

Hence, in the infinite $ N $ limit, the  (`pion') field $ \Phi_a\; (1
\leq a \leq N)$ is a free field with mass squared equal to  $ M^2 $.

On the other hand, the composite field $ {\vec \Phi}^2(x) $ has as
inverse propagator in euclidean 
space time given by 
\begin{equation}\label{Deltak}
\Delta(k) = \Delta(0) + {1 \over{(4\pi)^2}} \; \int_0^1 dx \; \log\left[1 +
x(1-x) {{k^2}\over {M^2}} \right] 
\end{equation}
where
\begin{equation}\label{Delta0}
\Delta(0) = -\frac2{\lambda} - {1\over{16\pi^2}} 
\log\left({{|m|^2}\over{e\; M^2}}\right)
\end{equation}

The relevance and properties of this propagator will be investigated below for
the cases of unbroken $m^2 >0$ and broken $m^2<0$ symmetry.

It is clear from the gap equation (\ref{gap}) and fig. (\ref{fig1}) , that $
\phi^2 $ as a function of $ M^2 $ has a maximum at
\begin{equation}
M^2 = M^2_0 \equiv |m|^2 \, 
\exp\left({{32\pi^2}\over \lambda} - 1 \right) \label{chi0}
\end{equation} 
for real $ M^2 $. Notice that such mass scale is of the order of the Landau
ghost. That means that it is outside the expected domain of validity of the $
\left({\vec \Phi}^2\right)^2 $, which can be used as an effective theory for
energies well below the Landau ghost.

At the maximum of $ \phi^2 = \phi^2(M^2) $, we have

\begin{equation}\label{ficero2}
\phi^2_0 \equiv \phi^2(M^2_0) = {{M^2_0}\over{16\pi^2}} - 
{{2\, m^2}\over{\lambda}} \; .
\end{equation} 
This value is also of the order of the Landau ghost for our choice of  $ m^2 $.

Therefore, for scales much lower than the Landau ghost, the $(\vec{\Phi}^2)^2$
theory is a sensible low energy effective description. In particular in our
studies of the dynamics the scale of the expectaction value is typically of
order $\phi^2 \approx |m|^2/\lambda << \phi^2_0$ which is well within the
regime of validity of the effective theory for weak coupling.

As observed in \cite{aks,barmo}, the inverse function $ M^2 = M^2(\phi^2) $ has
two branches as a function of $ \phi^2 $. However, the second branch, $ M^2 >
M^2_0 $, (branch II in \cite{aks}) corresponds to pion masses beyond the Landau
ghost.  A consistent low energy effective theory emerges for energies well
below $ M^2_0 $ [See ref\cite{barmo}, sec. IIIA]. This low energy condition
requires that this scalar field theory be studied with a finite but large
ultraviolet cutoff $ |m| << \Lambda << |m|\; e^{16\pi^2/\lambda} $ and a
positive coupling $ \lambda $.

\subsection{Analysis of the gap equation and the effective potential
for $m^2 > 0$}

Equation (\ref{gap}) and fig. 1 reveal that for $ m^2 > 0 $, the gap
equation has 
a real and positive solution for $ 0 \leq \phi^ 2 \leq \phi^2_0 , \; M^2_1 \leq
M^2 \leq M^2_0 $ where $ M^2_1 $ corresponds to $ \phi^2 = 0 $.  The field
$\phi $ becomes complex and hence unphysical for $ M^2 \leq M^2_1 $.

In fig. 2, the point
$$
w_1 \equiv  e^{-32\pi^2/\lambda} \; \frac{M^2_1}{m^2} \; .
$$
corresponds to the intersection of the lower curve (unbroken symmetry) with the
$ \varphi = 0 $ axis. $ w_1 $ is thus the positive solution of the
trascendental equation
$$
w_1 \log w_1 = -  \frac{32 \pi^2}{\lambda}  \;
e^{-\frac{32\pi^2}{\lambda}}  \; .
$$
[cfr. eq.(\ref{fiw})] which defines the value of $M^2_1$.

Clearly, the effective potential is minimal for $ \phi^2 = 0, \; M^2 = M^2_1
$. This is the $ O(N) $ symmetric ground state with pions of mass $ M^2_1 > 0$
\cite{barmo}.

The study of the field theory dynamics in this unbroken symmetry case for
initial excited states\cite{late} shows that the (time dependent) effective
mass tends to a nonzero value for late times. The pions are thus massive and
the zero mode slowly dissipates its energy (there is a non-zero threshold for
particle production) and attains the ground state ($ \phi = 0 $) very slowly,
with an oscillatory behavior modulated in amplitude by a power law with
anomalous relaxation exponent for asymptotically large time\cite{late}.

We see that the static analysis through the effective potential agrees with the
long time dynamics in the sense that the expectation value of the scalar field
relaxes to the minimum of the effective potential, i.e. $ \phi =0 $ with
oscillations that reflect a non-zero mass.  In addition, the effective pion
mass obtained asymptotically in the dynamical situation depends on the initial
energy density (which is conserved)\cite{late}. This dependence of course
cannot be captured by a static analysis which only applies to the ground state.

The $ {\vec \Phi}^2 $ propagator given by eqs.(\ref{Deltak}-\ref{Delta0}) has a
tachyonic pole at $ \Delta(m_T) = 0 $, if $ \Delta(0) < 0 $. This occurs for $
M^2 < m^2\; \exp\left({{32\pi^2}\over\lambda} - 2 \right) $. That is, a tachyon
is present unless the pion mass is of the order (or larger) than the Landau
ghost. In particular, tachyons are absent if one chooses the second branch of
the function $ M^2 = M^2(\phi^2) $ \cite{aks}.  As stated before, this scalar
theory should only be interpreted to be valid as a low energy effective theory
with $ \log\frac{M^2}{m^2} = O(1) $. The tachyon is then present but at a very
large mass provided $ \lambda $ is small. We find from
eqs.(\ref{Deltak}-\ref{Delta0}),
$$
\Delta(k) \buildrel{k \to \infty}\over= -\frac2{\lambda} + {1
\over{(8\pi)^2}}\log\left({k \over{m\, 
\sqrt{e}}}\right) + O({1 \over {k^2}})\; .
$$
Thus the tachyon mass 
$$
m_T = m \;\exp\left({{16\pi^2}\over \lambda} + \frac12 \right) \; .
$$
is beyond the range of validity of the effective $ \left({\vec \Phi}^2\right)^2
$ theory.

In particular in our studies of the dynamics the typical expectation values are
of order $\phi^2 \approx m^2/\lambda$ and for these energy scales ${\cal
O}(m/\sqrt{\lambda})$ the ultra-heavy tachyonic states are never excited and
they decouple from the low energy theory.  We can thus safely ignore the
existence of the tachyon.

\subsection{Analysis of the gap equation and the effective potential
for $ m^2 < 0 $}

In this case, it is clear from eq.(\ref{gap}) and fig. 2 that $ \phi^2 \geq
{{2\, |m|^2}\over {\lambda}} $ for $M^2$ real in the interval $ 0\leq M^2 \leq
M^2_0 $ with $M^2_0$ given by eq.(\ref{chi0}). Equation (\ref{potef}) and
fig. 2 show that the minimum of the effective potential corresponds to $ M^2 =
0, \; \phi^2 = {{2\, |m|^2}\over {\lambda}} $.  That is,
$$
\varphi^2(w)  \geq \varphi^2(0)= \frac{32 \pi^2}{\lambda }\;
e^{-\frac{32\pi^2}{\lambda}} 
$$
for real $ 0 \leq w \leq 1/e $. The minimum of $V_{eff} $ is at the point $ w =
0 $ in the upper curve (broken symmetry) in fig. 2.

This yields a {\em broken} $ O(N)$-symmetry ground state. As expected, we have
$ N - 1 $ massless pions which are the Goldstone bosons of the $ O(N) \to
O(N-1) $ breaking \cite{barmo} as can be seen from eq.(\ref{potef}) and fig. 2
which show that $ M^2 = 0 $ is the minimum of the effective potential for $ M^2
\geq 0 $ and below the Landau ghost.  However, this value corresponds to a
non-negative $ \phi^2 $ only for $m^2 < 0$. The fact that the minimum $ M^2 = 0
$ of the effective potential happens in the real configuration space (for $ m^2
< 0 $) is in agreement with the asymptotic late time behavior found in
refs.\cite{dis,linon,big,late}. There it was found that the dynamical (time
dependent) effective mass of the pion fields tends asymptotically to zero for $
m^2 < 0 $ (and initial energy below the false vacuum but above the ground
state).  The presence of massless particles allows the zero mode energy to
dissipate efficently into them and the zero mode rapidly reaches the ground
state. This happens very slowly in the $ m^2 > 0 $ case (unbroken symmetry). In
this case the pions are massive and energy dissipation occurs via non-linear
resonances\cite{late}.

Now we come to one of the important features of the effective potential in the
broken symmetry phase: it follows from eqs.(\ref{potv}),
(\ref{fiw}) and (\ref{ficero2}) that  there is no real solution $  M^2 $
(or  $w$) for $ \phi^2 > \phi^2_0 $. Therefore,  $  M^2 $ and $w$ as well as
the effective potential become {\em complex}  despite of the fact that
the field $ \phi^2 $ is real. 

Whereas in a loop 
expansion this imaginary part is typically ignored with the hope that higher
order corrections (loops) could remedy this situation, in the large $ N $ limit
this argument does not hold.  The physics associated with the {\em imaginary
part} of the effective potential has been clarified in
references\cite{erickwu,boyvega}.

In the region in which the effective potential computed from eq.(\ref{potef})
or eq.(\ref{potv}) turns to be complex, referred to as the spinodal region, the
system is in a mixed state in which several phases coexist. The true
equilibrium description of the system in the spinodal region is in terms of a
Maxwell constructed effective potential, which is flat in the coexistence
region $ 0 \leq \phi^2 \leq {{2\, m^2}\over {\lambda}} $\cite{rg}.  Although
such construction is rather arbitrary at the level of phenomenological
statistical mechanics (the lever rule), a consistent renormalization group
analysis has revealed that the infrared limit of a carefully coarse grained
effective action yields the Maxwell constructed effective
potential\cite{rg}. We also note that in this case of continuous symmetry the
spinodal region extends all the way from the maximum to the minimum of the
potential without a region of metastability. This must be contrasted with the
case of discrete symmetry, wherein the spinodal ends in the limit of
metastability {\em before} the minimum of the potential.  The fact that the
spinodal region reaches from the maximum to the minimum of the potential is a
consequence of the presence of Goldstone bosons.

At this point we summarize the relevant features of the effective potential in
the large $ N $ limit to compare to the results obtained from the dynamical
evolution\cite{dis,linon}.

\begin{itemize}
\item{ In the unbroken symmetry case with positive renormalized mass squared ($
m^2 > 0 $) the static effective potential has a unique minimum for a nonzero value of
the squared mass $ M^2 $ of the excitations and vanishing expectation value of
the field $ \phi $. The dynamics agrees with a vanishing expectation value of
the field reached asymptotically at late times with power laws\cite{late}. In
addition, the asymptotic effective mass obtained from the dynamical evolution
depends on the initial expectation value of the field\cite{late}. }

\item{The analysis of ref.\cite{barmo} shows that the $ O(N) $ symmetry is
broken in the effective theory for $ m^2 < 0 $ where again, by effective
theory we imply describing energies well below the Landau ghost scale $ M^2_0 $
[see eq.(\ref{chi0})].

When the renormalized mass $ m^2 < 0 $, the minimum of the static effective potential
is found at $ M^2 = 0 $ and for a nonzero value of the field $ \phi $. The $
O(N) $ symmetry is then spontaneously broken and the massless scalars are the
corresponding Goldstone bosons. The static effective potential computed from
eq.(\ref{potef}) or eq.(\ref{potv}) is complex in the interval $ 0 < \phi^2 <
2|m|^2/\lambda $; this is the spinodal region that signals phase
coexistence. The spinodal region reaches from the maximum to the minimum of the
potential unlike in the discrete symmetry case. 

The physics inside the spinodal region cannot be captured via a static
description.  An {\em equilibrium} description of this region can be given in
terms of the Maxwell construction which leads to a flat effective potential for
$ 0< \phi^2 < 2 m^2/\lambda $. Such a description emerges in quantum field
theory with a consistent treatment using the {\em exact} renormalization
group\cite{rg,polo,tetradis}.  In the next sections 
we will study the {\em dynamical} Maxwell construction that leads to a
flat  {\em dynamical} effective potential in the spinodal region via the
non-equilibrium dynamics. 

In section IV we study the dynamics of evolution in a broken symmetry case
but when the initial value of $\phi$ is {\em above} the maximum of the
tree level potential. In this case the dynamics is similar to that of the
unbroken symmetry case. The effective potential is simply unable to describe
correctly this physical situation. }

\item{Furthermore, the effective potential clearly illuminates the regime of
validity of the scalar theory as an effective, cutoff low energy theory below
the Landau ghost.  Obviously, the physics is radically different when one
considers pions and/or field expectations values at the energy scale of the
Landau ghost. Our study of the dynamics in both the broken and unbroken
symmetry phases will be restricted to energy scales $\approx
|m|/\sqrt{\lambda}$ and energy densities $\approx m^4 /\lambda$ i.e. well below
the Landau ghost scale for weak coupling. Therefore this scalar theory is a
reliable cutoff low energy model to study the non-perturbative dynamics for
energy scales below the Landau ghost.}
\end{itemize}

\section{Non-Equilibrium Dynamics in the large $ N $ limit: The Density Matrix}

The field theory tool to study non-equilibrim dynamics is the quantum density
matrix $\hat{\rho}(t)$ which obeys the functional Liouville equation
$$
i \frac{\partial \hat{\rho}}{\partial t} = \left[H(t),\hat{\rho}\right]
$$
where we admit a Hamiltonian that can depend explicitly on time as it
would be the case in expanding cosmologies. A particularly illuminating
representation for the field theoretical density matrix is the Schr\"odinger
representation\cite{FRW,mottola}, in which the field operator is
diagonal. In such a 
representation
$$
\rho[\Phi(\cdot),\tilde{\Phi}(\cdot);t] = \langle \Phi | \hat{\rho}(t)|
\tilde{\Phi} \rangle
$$
and the diagonal density matrix elements
$\rho[\Phi(\cdot),{\Phi}(\cdot);t]= {\cal{P}}(\Phi;t)$ give the
functional probability density for finding a field configuration with profile
$\Phi(\vec x)$ at time t. Therefore, this representation allows to ask
questions about the formation of semiclassical configurations in terms of
probability functionals. This is the correct interpretation for the question of
finding {\em any} arbitrary field configuration in the ensemble described by
the non-equilibrium density matrix.

We will study the general case of a biased initial condition leading to
symmetry breaking dynamics to illuminate various features of the
non-equilibrium dynamics and the implementation of Goldstone's
theorem. Therefore we write
\begin{eqnarray}
\vec{\Phi}(\vec x,t) & = &  \left( \sqrt{N} \phi(t)+\sigma(\vec x,t),
\vec{\Psi}(\vec x,t) \right) \nonumber \\ 
\langle \Phi^a(\vec x,t) \rangle & = & \sqrt{N}\; \phi(t)\;\delta^{a,1}  =  
Tr\left[\hat{\rho}(t) \Phi^a(\vec x)\right] \nonumber
\end{eqnarray}
and $\vec{\Psi}$ is an $ O(N-1) $ vector.  

The canonical momenta conjugate to $\sigma$ and $\Psi^a$ are given by
$$
\Pi_{\sigma}(\vec{x},t)  = \dot{\sigma}(\vec{x},t) \; \; ; \; \; 
\Pi^a_{\Psi}(\vec{x},t) = \dot{\Psi}^a(\vec{x},t)
$$
and the Hamiltonian becomes
$$
H(t) = \int d^3x \left\{ \frac{1}{2}{\Pi^2_{\sigma}}+
\frac{1}{2}\vec{\Pi}^2_{\Psi}+
\frac{1}{2}(\vec{\nabla}\Phi)^2+\frac{1}{2}(\vec{\nabla}\vec{\Psi})^2+
V(\phi,\sigma,\vec{\Psi}) \right\}+ \frac{N}{2} \dot{\phi}^2(t)
$$

In the Schr\"{o}dinger representation (at an arbitrary fixed time $t_0$), the
canonical momenta are represented as functional derivatives
\begin{eqnarray}
 \Pi_{\sigma}(\vec{x}) & = &  -i \frac{\delta}{\delta \sigma(\vec{x})} 
\nonumber \\
 \Pi^a_{\Psi}(\vec{x}) & = &  -i \frac{\delta}{\delta \Psi^a(\vec{x})}
\nonumber
\end{eqnarray}
and the Liouville equation for the
density matrix becomes a functional differential equation\cite{FRW}
\begin{equation}
i  <\Psi,\sigma |\dot{\rho}(t)| \tilde{\sigma},\tilde{\Psi}> =
\left(H[\Pi_{\sigma},\vec{\Pi}_{\Psi}, \sigma,\vec{\Psi};t]-
H[\tilde{\Pi}_{\sigma}, 
\vec{\tilde{\Pi}}_{\Psi}, \tilde{\sigma},\vec{\tilde{\Psi}};t]\right)
<\Psi,\sigma |\rho(t)| \tilde{\sigma},\tilde{\Psi}> \label{liouvilleqn} .
\end{equation}

The large $ N $ limit can be implemented by introducing an auxiliary field or
alternatively using the following factorization

\begin{eqnarray}
\sigma^4 & \rightarrow & 6 \langle \sigma^2 \rangle \sigma^2 +\mbox{ constant }
\label{hartreechi4} \\
\sigma^3 & \rightarrow & 3 \langle \sigma^2 \rangle \sigma 
\label{hartreechi3} \\
(\vec{\Psi}\cdot\vec{\Psi})^2
							& \rightarrow
							& 2\langle
							\vec{\Psi}^2
							\rangle 
\vec{\Psi}^2 - \langle \vec{\Psi}^2 \rangle^2 \label{hartreepi4} \\
\vec{\Psi}^2 \sigma^2
							& \rightarrow
							& \vec{\Psi}^2
							\langle
							\sigma^2
							\rangle+
							\langle  
\vec{\Psi}^2 \rangle \sigma^2 \label{hartreepi2chi2} \\
\vec{\Psi}^2 \sigma
							& \rightarrow
							& \langle
							\vec{\Psi}^2
							\rangle \sigma
							, 
\label{hartreepi2chi}
\end{eqnarray}   
where the constant term in (\ref{hartreechi4}) is irrelevant as it will give a
subleading correction in the large $ N $ limit.  In this approximation, the
potential (\ref{barepotential}) becomes
\begin{eqnarray}
V(\phi,\sigma,\vec{\Psi})  & = &  V(\sqrt{N}\phi)+
\sigma \; {\cal{V}}^1(t)+
\frac{M^2_{\sigma}(t)}{2}\;
\sigma^2+\frac{M^2(t)}{2}\;\vec{\Psi}^2 \nonumber\\
 V(\sqrt{N}\phi) & = & {N \, \lambda \over 8}\left( \phi^2 - {2 \, m^2
\over \lambda} \right)^2 \label{potV}\\
{\cal{V}}^1(t)       & = & \sqrt{N}\phi(t) \left[m^2+\frac{\lambda}{2}
\phi^2(t)+ 
\frac{\lambda}{2N} \langle \vec{\Psi}^2 \rangle(t) \right] \label{nu1} \\
{M}^2(t) & = & m^2+\frac{\lambda}{2} \phi^2(t)+
\frac{\lambda}{2N}
\langle \vec{\Psi}^2 \rangle (t) \label{pionmass} \\
{M}^2_{\sigma}(t)& = & m^2+\frac{3\lambda}{2} \phi^2(t)+
\frac{\lambda}{2N}
\langle \vec{\Psi}^2 \rangle (t)  \label{chimass}
\end{eqnarray}

Since in this approximation the evolution Hamiltonian is quadratic (at the
expense of a self-consistency condition), and the different $ O(N) $ components
evolve independently, the density matrix factorizes as\cite{FRW}

\begin{equation}
\rho(t) = \rho_{\sigma}(t)\otimes \rho_{\Psi}(t) \nonumber
\end{equation}

Furthermore, the density matrices can be chosen to be Gaussian, and will remain
Gaussian under time evolution.

Using spatial translational invariance, we can decompose the fields into their
spatial Fourier modes and use these as a basis for the Schr\"{o}dinger
representation density matrix:

\begin{eqnarray}
\sigma(\vec{x},t) & = & \int\frac{d^3k}{(2\pi)^{3/2}}\;
\sigma_{\vec{k}}(t)  \; \exp(-i\vec{k} \cdot \vec{x}) \\
\vec{\Psi}(\vec{x},t)    & = & \int\frac{d^3k}{(2\pi)^{3/2}}\;
\vec{\Psi}_{\vec{k}}(t) \; \exp(-i\vec{k} \cdot \vec{x}) \; .
\end{eqnarray}

The Gaussian density matrices acquire a simple form in terms of these
Fourier modes\cite{FRW} 

\begin{eqnarray}
\rho_{\sigma}[\sigma,\tilde{\sigma},t] & = & \prod_{\vec{k}} 
{\cal{N}}_{\sigma,k}(t) \; 
\exp\left\{- \frac{A_{\sigma,k}(t)}{2}\; \sigma_k(t)\sigma_{-k}(t)-
\frac{A^*_{\sigma,k}(t)}{2}\; \tilde{\sigma}_k(t)\tilde{\sigma}_{-k}(t)
\right. 
\nonumber \\
& -  & \left. 	B_{\sigma,k}(t)\sigma_k(t)\tilde{\sigma}_{-k}(t) 
+i p_{\sigma,_k}(t)\left[\sigma_{-k}(t)-
\tilde{\sigma}_{-k}(t)\right] \right\}\; , \label{chidensitymatrix} 
\nonumber \\
\rho_{\vec{\Psi}}[\vec{\Psi},\tilde{\vec{\Psi}},t] & = &
\prod_{\vec{k}} {\cal{N}}_{\Psi,k}(t) 
\exp\left\{- \frac{A_{\Psi,k}(t)}{2}\; \vec{\Psi}_k\cdot \vec{\Psi}_{-k}-
\frac{A^*_{\Psi,k}(t)}{2}\; \tilde{\vec{\Psi}}_k\cdot
\vec{\tilde{\Psi}}_{-k} \right. 
\nonumber \\ & -  &  \left. B_{\Psi,k}(t)\; \vec{\Psi}_k\cdot 
\tilde{\vec{\Psi}}_{-k}\right\}\;  . \label{psidensitymatrix} 
\end{eqnarray}

Only the component along the broken symmetry direction, $\sigma$ has an
expectation value of the momentum, represented by $p_{\sigma,k}(t)$ in
(\ref{chidensitymatrix}), whereas the transverse directions have vanishing
expectation value of the momentum. The kernels $A_{\Psi}; \; B_{\Psi}$ are
scalars under $ O(N-1) $.  Although we will ultimately work at zero
temperature, we have written the most general form of the density matrix,
including a mixing term given by the kernels $B_{\sigma,\Psi,k}$ above to
display the most general form of the density matrix. When the mixing kernels
$B_{\sigma,\Psi,k}$ vanish, the density matrix collapses to a product of a wave
functional and its complex conjugate, i.e. it describes a pure state. Also note
that hermiticity of the density matrix requires that the mixing kernels
$B_{\sigma,\Psi}$ be real.

Now the Liouville equation (\ref{liouvilleqn}) can be solved for the time
dependent kernels by comparing the different powers of $\sigma, \Psi$ on both
sides of the equation (\ref{liouvilleqn}). We obtain the following set of
evolution equations for the kernels\cite{FRW}:

\begin{eqnarray}
i\,\frac{{\dot{\cal{N}}}_{\sigma,k}}{{\cal{N}}_{\sigma,k}} & = &
\frac{1}{2}(A_{\sigma,k}-A^*_{\sigma,k}) \label{normchi} \\
i\dot{A}_{\sigma,k}  & = & 
A^2_{\sigma,k}-B^2_{\sigma,k}-\omega^2_{\sigma,k}(t)
	\label{achi} \\
i\dot{B}_{\sigma,k}   & = &  B_{\sigma,k}
\left(A_{\sigma,k}-A^*_{\sigma,k}\right) \label{bchi} \\
\omega^2_{\sigma,k}(t)                                 & = & 
k^2+M^2_{\sigma}(t) \label{omega2chi} \\
\ddot\phi+   M^2(t)\phi & =&  0 \; . 
\label{fieqnofmotion} \\
i\,\frac{{\dot{\cal{N}}}_{\Psi,k}}{{\cal{N}}_{\Psi,k}}  & = &
\frac{1}{2}(A_{\Psi,k}-A^*_{\Psi,k}) \label{normpi} \\
i\dot{A}_{\Psi,k}                                    & = & 
A^2_{\Psi,k}-B^2_{\Psi,k}-\omega^2_{k}(t) 	\label{api} \\
i\dot{B}_{\Psi,k}                                    & = & 
B_{\Psi,k}\left(A_{\Psi,k}-A^*_{\Psi,k}\right) \label{bpi} \\
\omega^2_{k}(t)                                 & = & 
k^2+M^2(t). \label{omega2pi}
\end{eqnarray}

Writing the kernels $A_k$ for $\sigma, \; \Psi$ in terms of real and imaginary
components $A_k(t) = A_{Rk}(t)+i A_{Ik}(t)$ and using the reality of the $B_k$,
we find the following invariants of evolution
\begin{equation}
\frac{B_k(t)}{A_{Rk}(t)} = \frac{B_k(t_o)}{A_{Rk}(t_o)}
\label{invar}
\end{equation}
 and 
\begin{equation}
\frac{{\cal{N}}_k(t)}{\sqrt{A_{Rk}(t)+B_k(t)}} = \mbox{constant}
\label{unitarity}
\end{equation}

The condition (\ref{unitarity}) reflects unitary time evolution of the density
matrix, i.e. the normalization is time independent.

At this point we notice that the dynamics of $\sigma$ decouples from that of
the rest of the fields i.e. it does not contribute to the evolution of the
order parameter or to that of the fluctuations of the $\Psi$ fields (transverse
directions). This is of course a consequence of the large $ N $ limit in which
the dynamics is solely determined by the $N-1$ transverse degrees of
freedom. 

At this point we trace out the density matrix of $\sigma$ and study the
evolution solely in terms of the order parameter $\phi$ and the density matrix
for the transverse directions.

The equations for the kernels can be simplified by 
defining\cite{FRW}
\begin{eqnarray}
A_{Rk}(t) & = &  {\cal{A}}_{Rk}(t) \coth[\Theta_k] \label{reala} \\
B_k(t) & = & -\frac{{\cal{A}}_{Rk}(t)}{\sinh[\Theta_k]} \label{boft}
\end{eqnarray}
with $ \Theta_k $ an arbitrary real parameter, which for an initial density
matrix in thermal equilibrium will be identified with the usual finite
temperature factor $\Theta_k = \omega_k/T$.
The complex quantity
\begin{equation}
{\cal{A}}_k(t) = {\cal{A}}_{R,k}(t)+iA_{Ik}(t)\label{acomplex}
\end{equation}
obeys a simple Ricatti differential equation
\begin{equation}
i\dot{{\cal{A}}}_k(t)= {\cal{A}}^2_{k}(t)-
\omega^2_{k}(t)  \label{ricattieq}
\end{equation}
which can be solved by introducing a complex field 
\begin{eqnarray}
&&{\cal{A}}_{k}(t)  =  -i \; \frac{\dot{\varphi}^*_k(t)}{\varphi^*_k(t)}
\label{ricasol} \\
&& \ddot{\varphi}^*_k(t)+\omega^2_{k}(t)\;  \varphi^*_k(t)  =  0. \label{phieq}
\end{eqnarray} 
The initial conditions on the mode functions are determined by the initial
values of the real and imaginary parts of the kernel $A_k(t=t_0)$. Since an
arbitrary time independent phase can be removed from the wave functionals, the
imaginary part of the kernel $A_k(t)$ at the initial time can be set to zero
without loss of generality.  Thus we parametrize the initial conditions on the
mode functions $\varphi_k(t)$ in the form
\begin{equation}
\varphi_k(t_0) = \frac{1}{\sqrt{2W_k}} \; \; ; \; \;
\left. \dot{\varphi}_k(t)\right|_{t_0} =  
-i \sqrt{W_k \over 2}\label{inicondmod}
\end{equation}
With this choice the initial density matrix for zero mixing kernels corresponds
to the product of vacuum wavefunctionals of harmonic oscillators of frequencies
$W_k$ and their complex conjugates, or if the mixing is of the thermal form it
corresponds to the density matrix of a thermal ensemble of harmonic oscillators
of these frequencies.

We find in the large $ N $ limit
\begin{equation}
\langle \vec{\Psi}^2(t)\rangle = N \int
\frac{d^3k}{(2\pi)^3}{|\varphi_k(t)|^2}  
\coth\frac{\Theta_k}{2} \label{fluc}
\end{equation} 
The mode functions $\varphi_k(t)$ form a complete set of states allowing us to
expand the {\em Heisenberg} operator for the transverse fluctuations as
\begin{equation}
 \vec{\Psi}(x) = \int\frac{d^3k}{(2\pi)^3}
\left[{\vec a}_k \; \varphi_k(t) \; e^{i\vec{k}\cdot \vec x} + {\vec a}^{\dagger}_k
 \; \varphi^*_k(t) \; e^{-i\vec{k}\cdot \vec x} \right], .
\end{equation}
where the annihilation and creation operators will allow us to make contact
with the Fock representation. With the initial conditions on the mode functions
(\ref{inicondmod}) we see that at the initial time $t_0$, to be chosen $t_0=0$
in what follows, the operators $\vec{a}^{\dagger}_k$ create quanta of frequency
$W_k$ out of the Fock vacuum state.

It is convenient and illuminating to introduce the {\em adiabatic} creation and
annihilation operators $\alpha_k(t)\; ; \; \alpha^{\dagger}_k(t)$ and expand
the Heisenberg field in the form

\begin{equation}
 \vec{\Psi}(x) = \int\frac{d^3k}{(2\pi)^3} \left[{\vec \alpha}_k(t) \;
\frac{e^{-i\int dt' \omega_{k}(t')}}{\sqrt{2\omega_{k}(t)}} \; e^{i\vec{k}\cdot
\vec x} + {\vec \alpha}^{\dagger}_k(t) \; \frac{e^{i\int dt'
\omega_{k}(t')}}{\sqrt{2\omega_{k}(t)}} \; e^{-i\vec{k}\cdot \vec x} \right]
\end{equation} 
The adiabatic operators diagonalize the instantaneous Hamiltonian and will
provide a clear description of the particle production contribution to the
total energy. The adiabatic operators are related by a Bogoliubov
transformation to the operators $a_k \; ; \; a^{\dagger}_k$ that create
excitations with respect to the Fock basis at the initial time. These operators
are available whenever the time dependent frequencies are real.

We now collect the final set of equations that determine the full dynamics of
the density matrix in the large $ N $ limit

\begin{eqnarray}
&& \ddot{\varphi}_k(t)+ \left[ k^2 +
m^2+\frac{\lambda}{2}\phi^2+\frac{\lambda}{2N} \langle \vec{\Psi}^2(t)
\rangle \right] \varphi_k(t)  =  0 \label{modeq} \\ 
&& \ddot\phi+\left[m^2+\frac{\lambda}{2}\phi^2+\frac{\lambda}{2N}
\langle \vec{\Psi}^2(t) \rangle \right] \phi(t)  =  0
\label{zeromodeqn} \\ 
&& \langle \vec{\Psi}^2(t) \rangle  =  \int \frac{d^3k}{(2\pi)^3}
{|\varphi_k(t)|^2}  \coth\frac{\Theta_k}{2}
\end{eqnarray}

These equations of motion are the same as those obtained in the Heisenberg
representation in the large $ N $ limit. The renormalization aspects have been
studied in references\cite{big,baacke1,late,losa,baacke2,FRW} to which we refer
the reader for details.

Although we have obtained the equations of motion in the general
case of a mixed density matrix, we will consider in what follows the
situation of an initial pure state by setting $\Theta_k=0$.

\section{Dynamics for broken symmetry potential }
In this section we study in detail the dynamics in a broken symmetry
potential $m^2 <0$ in two important cases: 
i) $  {\cal V}(\eta_0) <
 {\cal V}(0) $ corresponding to $ |\eta_0| < \sqrt{2}$, ii)  $  {\cal
V}(\eta_0) >  {\cal V}(0) $ corresponding to $ |\eta_0| > \sqrt{2} $, where we
define 
\begin{eqnarray}
V(\sqrt{N}\phi)& =& \frac{2 m^4 N}{\lambda}~ {\cal{V}}(\eta)\nonumber \\
{\cal{V}}(\eta) & = & \frac{1}{4} (\eta^2-1)^2 \label{defVeta}.
\end{eqnarray}
We will analyze both of these situations numerically below. Our results can be
summarized as follows.

In the first case, $ {\cal V}(\eta_0) < {\cal V}(0) $, the energy is such that dynamics 
corresponds to a
spontaneously broken situation in the sense that if $\eta_0 \neq 0$ the
dynamical evolution results in a non-zero asymptotic value for
$\eta(\infty)$. In these cases the effective mass of the transverse
fluctuations vanishes asymptotically, i.e. these fluctuations are Goldstone
bosons\cite{big,late}.  As it will become clear from the dynamics, in this
situation the Maxwell construction is realized through the non-perturbative
production of Goldstone bosons.
 
In the second case, $ {\cal V}(\eta_0) > {\cal V}(0) $, the energy is large enough that under the dynamical
evolution the expectation value samples the minima ergodically. Its asymptotic
value oscillates around $\eta =0$ with diminishing amplitude and with a
non-zero asymptotic effective mass for the fluctuations despite the fact that
$m^2<0$. In this case particle production is a consequence of parametric
amplification.

\subsection{Equations of Motion: Numerical Analysis}

Here we set up the form of the equations of motion for the expectation value
and the relevant modes in a form amenable to numerical analysis.

It proves convenient to introduce the following dimensionless quantities
in terms of the renormalized mass and couplings
(\ref{massrenor},\ref{coupren}): 
\begin{eqnarray}
&&\overline{|m|}^2  =   |m|^2- 
\frac{\lambda}{2N}  \langle \Psi^2({\vec x},t=0)\rangle_R  \; \; , \; \;
\frac{1}{\overline{\lambda}}  =   \frac{1}{\lambda}+
\frac{1}{32\pi^2} \label{overlinem}\\ 
&& \tau = \overline{|m|}\, t \; \; , \; \;  q = \frac{k}{\overline{|m|} }
\; \; , \; \; \Omega_q= \frac{W_k}{\overline{|m|}} \; \; ,\; \;
\eta^2(\tau) = \frac{\overline{\lambda}}{2 \overline{|m|}^2 }\;
\phi^2(t)  \; \; ,  
\label{etadef} \\ 
&& g \Sigma(\tau) = 
\frac{\overline{\lambda}}{2\overline{|m|}^2 } \left[ \langle \Psi^2({\vec x},t) \rangle_R- \langle
 \Psi^2({\vec x},0) \rangle_R  
\right]  \; \; , \; \; \left( \; \Sigma(0) = 0  \;\right) \label{sigma} \\
&& g = \frac{\lambda}{8\pi^2}  \; \; , \; \;
U_q(\tau) \equiv \sqrt{\overline{|m|}} \; \varphi_k(t)  \; \; .\label{gren}
\end{eqnarray}
 Here  $ \langle \Psi^2({\vec x},t) \rangle_R $ stands
for the renormalized   
composite operator [see eq.(\ref{sigmafin}) for an explicit expression].

In the case of broken symmetry ( $ m^2 < 0 $) the field equations in the
$N = \infty$ limit become\cite{dis}:

\begin{eqnarray}\label{modo0R}
& & \left[\;\frac{d^2}{d\tau^2}+{\cal M}^2(\tau)\;\right]\eta(\tau)
  = 0  \label{expvaleqn} \\
& & \left[\;\frac{d^2}{d\tau^2}+q^2+{\cal M}^2(\tau)\;\right]
 U_q(\tau) =0 \label{modokR}
\end{eqnarray}
Here,
\begin{equation}\label{masef}
{\cal M}^2(\tau) \equiv -1+\;\eta^2(\tau) + g\;  \Sigma(\tau) 
\end{equation}
plays the r\^ole of a (time dependent) renormalized effective mass squared
and $ \Sigma(\tau)$ is given in terms of the mode functions $U_q(\tau)$
by\cite{dis,big,late} 
\begin{eqnarray}
g \Sigma(\tau) & = & g \int_0^{\infty} q^2
dq \left\{ 
\mid U_q(\tau) \mid^2 - \frac{1}{\Omega_q} 
+ \frac{\theta(q-1)}{2q^3}\left[ 
 -\eta^2_0 + \eta^2(\tau) + g \; \Sigma(\tau) \right] \right\} \;
. \label{sigmafin} 
\end{eqnarray}

The choice of boundary conditions is subtle for broken symmetry and a
distinction must be made for the cases in which the initial expectation value
of the scalar field is such that $\eta_0 >1$ or $\eta_0 <1$. In the first case,
the initial frequencies are all real and vacuum type initial conditions can be
chosen. In the second case, there is a band of unstable modes and the initial
conditions must be chosen differently. The resulting dynamics in weak coupling
is not very sensitive to the different choices of initial conditions in the
second case\cite{big}.  We shall use the following initial conditions for the
mode functions\cite{big,late}: for $ \eta_0> 1$, we choose
\begin{equation}
U_q(0) = {1 \over {\sqrt{ \Omega_q}}} \quad , \quad 
{\dot U}_q(0) = - i \; \sqrt{ \Omega_q}\; \; ; \; \; 
\Omega_q=  \sqrt{q^2 -1 + \eta^2_0} \label{conds12}
\end{equation}
while for $|\eta_0|<1$
\begin{eqnarray}
U_q(0) & = & {1 \over {\sqrt{ \Omega_q}}} \quad , \quad 
{\dot U}_q(0) = - i \; \sqrt{ \Omega_q}  \nonumber \\
\Omega_q & = & \sqrt{q^2 +1 +\eta^2_0} \quad {\rm for} \; \; q^2 <
q_u^2 \equiv 1-\eta^2_0 \label{unsfrequ} \\
\Omega_q & = & \sqrt{q^2 -1 + \eta^2_0}  \quad {\rm for} \;  \; q^2
>q_u^2 \quad  ; 
 \quad 0 \leq \eta^2_0 <1 \label{stafrequ} \; .
\end{eqnarray}
To these we add initial conditions for the zero mode given by 
$$
\eta(0) = \eta_0  \quad , \quad {\dot\eta}(0) = 0 
$$

Although the notion of particle is not unique in a time dependent
non-equilibrium situation, a suitable definition can be given with respect to
some particular pointer state. Two definitions of the particle number prove
particularly important for the discussion that follows. The first is the
particle number defined with respect to the initial Fock vacuum state. This
takes the following form in terms of the dimensionless mode functions we find
the number of particles of (dimensionless) momentum $q$ per unit volume to be
given by\cite{big}
\begin{eqnarray}
N_q(\tau) &=& \frac{1}{4}\left[
\Omega_q |U_q(\tau)|^2+\frac{|\dot{U}_q(\tau)|^2}{\Omega_q}
\right] -\frac{1}{2}\; . \label{partnumber}
\end{eqnarray}

For the stable modes, i.e. for those for which the adiabatic frequencies are
real at all times we can also consider the adiabatic particle number. It is
given by\cite{late,mottola}
\begin{eqnarray}
N^{ad}_q(\tau) &=& \langle \alpha^{\dagger}_q(\tau) \alpha_q(\tau) \rangle  
 \label{adpart} \\
&=& 
\frac{1}{4}\left[ \omega_q(\tau) \;
|U_q(\tau)|^2+\frac{|\dot{U}_q(\tau)|^2}{\omega_q(\tau)} 
\right] -\frac{1}{2}\; , \nonumber
\end{eqnarray}
where $  \omega_q(\tau) = \sqrt{ q^2 +  {\cal M}^2(\tau)} $.
Both definitions coincide at $ \tau = 0 $.

We choose a initial state with zero temperature so the the initial occupation
number $ N_q(0) $ vanishes. It is easy to generalize our results to the case of
initial states with a thermal distribution.

\subsection{ $ {\cal V}(\eta_0) <  {\cal V}(0) ,\;  ( |\eta_0| < \sqrt2 ) $:
Evolution towards spontaneously broken symmetry states}

In this case the generic features of the evolution were studied in
refs.\cite{big,late}. If initially $\eta_0 = \dot{\eta}_0=0$, then $\eta$
remains at zero; this is a fixed
point of the evolution equations. For early times, the mode functions
grow exponentially as a consequence of spinodal instabilities and
their amplitude become non-perturbatively large $|U_q(\tau)|^2 \approx 1/g $
for $\tau > \tau_s \approx \ln 1/g $ (see eq.(\ref{spinotime}) below and
references\cite{big,late}). For $\eta_0 \neq 0$ we find that, whereas the
effective mass of the fluctuations vanishes asymptotically leading to
Goldstone bosons, the expectation value $ \eta(\tau) $ tends
asymptotically to a 
constant $ \eta_{\infty} $ that depends on the initial conditions. Fig. 3
displays this evolution for two different initial expectation values.  It
clearly shows that $ \eta_{\infty} $ depends on the initial
condition. Furthermore we can see from this figure that the effective frequency
of oscillation becomes smaller at larger times, consistent with a vanishing
effective mass and the production of Goldstone bosons (see Fig. 5).

The relation between the asymptotic value $\eta_{\infty}$ and the initial value
$\eta_0$ is shown in fig. 4 for $ 0 < \eta_0 < \sqrt2 $.  We find that $
\eta_{\infty} $ vs. $ \eta_0 $ can be fitted remarkably well (see fig. 4) by
the power law behavior
\begin{equation}\label{leyx}
\eta_{\infty}(\eta_0) = \left[  \eta_0^2 \, (2 -  \eta_0^2) \right]^x
\;  = ( 1 - 4 \, E)^x \; ,
\end{equation}
with $x$ a dynamical critical exponent. A numerical fit yields $ x = 0.25\ldots
$ for small coupling. This exponent is rather insensitive to the coupling for
weak coupling.  Analogous dynamical exponents were found in ref.\cite{late} for
the late time evolution of the expectation value in the {\em unbroken} symmetry
case.

The law (\ref{leyx}) is invariant under the exchange $ \eta_0 \Leftrightarrow
\sqrt{2 - \eta_0^2} $ showing that $ \eta_{\infty}(\eta_0) $ is solely a
function of the total initial (dimensionless) energy $ E = \frac14
(\eta_0^2-1)^2 $. In addition, we have $ \eta_{\infty}(1) = 1 $ and $
\eta_{\infty}(0) = 0 $, as expected, since these are fixed points of the
dynamics for $\dot{\eta}(t=0)=0$. Furthermore we find that the effective time
dependent mass given by (\ref{masef}) vanishes asymptotically as\cite{late}
\begin{equation}
{\cal M}^2(\tau) \buildrel{\tau \to \infty}\over=
\frac{A}{\tau} \cos\left[2\tau+2 \, c\,
\ln\frac{\tau}{\tau_s}+\gamma\right] \label{latemass} 
\end{equation}
\noindent with $A\; ; \; c$ given in reference\cite{late}. Figure 5
shows $ {\cal M}^2(\tau) $. This asymptotic form of the effectiva mass
translates into the following asymptotic behavior of $\eta(\tau)$
\begin{equation}
\eta(\tau)\buildrel{\tau \to \infty}\over=
\eta_{\infty}(\eta_0)\left[1+\frac14 {\cal M}^2(\tau)\right]+{\cal
O}(1/\tau^2) \label{asiformula} 
\end{equation}

\subsection{Dynamical Maxwell construction:}
The results of the numerical analysis lead to two important
consequences that must be highlighted: 
\begin{itemize}
\item{For $|\eta_0| < \sqrt{2}$ {\em all} values of $0\leq
|\eta_{\infty}|\leq 1$ can be reached asymptotically 
depending on the initial expectation value (energy). This is
manifest in the expression  (\ref{leyx}) for the asymptotic expectation
value and is the dynamical equivalent of the `lever rule' for the
Maxwell construction in equilibrium statistical mechanics.   }

\item{Asymptotically the effective mass squared, i.e. the
``restoring'' force in the evolution  of $\eta(\tau)$ 
vanishes as given by eq. (\ref{latemass}) (see fig. 5), compatible
with massless Goldstone bosons.  } 

\item{The states for which $0\leq |\eta_{\infty}|< 1$ are excited states
with large energy densities,  the smaller $\eta_{\infty}$ the larger the
energy density. Most of the energy is stored in long-wavelength condensates
as a result of non-perturbative particle production via spinodal
instabilities.} 

\end{itemize} 

These  features of the dynamics combined lead to the conclusion
that all of the broken symmetry states 
with asymptotic equilibrium values $0\leq |\eta_{\infty}|\leq 1$ are
allowed and that the dynamical potential associated 
with these states is {\em flat} since the ``restoring force'' for  the
evolution of the order parameter is ${\cal M}^2(\tau)$ which vanishes asymptotically as a result of Goldstone's theorem. Therefore we see the emergence of
a {\em dynamical} Maxwell construction. 
The energy initially stored in one mode, (the expectation value or
zero mode) is transferred  
to all of the other modes but mainly to those that are spinodally
unstable. The   amplitude of these modes  becomes non-perturbatively 
large $\sim 1/g$. 

This should be contrasted with the results from the static effective
potential that predicts that the {\em only} equilibrium 
broken symmetry solution which is physically acceptable is $|\eta|
=1$, states with $0\leq |\eta_0| \leq 1$ lead to  
an imaginary part, which as described previously is the hallmark of
the spinodal instabilities leading to non-perturbative particle
production.  

In fact, we can make contact with the static effective potential picture in the following way.

Since we are considering translationally and rotationally invariant states, the
expectation value of $ T^{\mu \nu} $ takes the perfect fluid form.  The energy
density, i.e. expectation value of the Hamiltonian (divided by the volume and
N) in the broken symmetry case is given by\cite{big}
\begin{eqnarray}
\varepsilon = \frac{\langle H \rangle}{NV}= \frac{\dot{\phi}^2}{2}
-\frac{|m_0|^2}{2} \; \phi^2 + \frac{\lambda_0}{8}\;\phi^4 -
\frac{\lambda_0}{8N} \;
\langle \vec{\Psi}^2 \rangle^2 + \frac{1}{4\pi^2}\int k^2\; dk \;\left[
|\dot{\varphi}^2_k(t)| + \omega^2_k(t) |\varphi_k(t)|^2\right]
\label{ener} 
\end{eqnarray}

In ref.\cite{big} it was proven that after a constant subtraction $\propto
\Lambda^4$ the energy is {\em finite }, furthermore (\ref{ener}) is {\em
conserved} by use of the eqs.(\ref{modo0R}) and (\ref{modokR})\cite{big}.

Using eq.(\ref{adpart}) and eq.(\ref{sigmafin}) we write the last term
in (\ref{ener}) in the following form
\begin{eqnarray} 
& \frac{1}{4\pi^2}&\int_0^{\Lambda}\;
k^2\;dk\;\left[|\dot{\varphi}_k(t)|^2+\omega^2_k(t)|\varphi_k(t)|^2
\right] = 
\varepsilon_U+  \frac{1}{4\pi^2}\int_{k_u}^{\Lambda}\;k^2\;dk\;
\omega_k(t)\left({N}^{ad}_k(t)+\frac{1}{2}\right)\; ,
\label{enefluc} \\
&\varepsilon_U &\equiv \frac{1}{2}\int^{k_u}_0 \frac{k^2\;dk}{2\pi^2}\;
\left[|\dot{\varphi}_k(t)|^2+\omega^2_k(t) \;
|\varphi_k(t)|^2 \right] \label{unstener}
\end{eqnarray}
where $k^2_u$ is the maximum unstable frequency in the broken symmetry case for
the case $\eta^2_0 <1$. The term $\varepsilon_U$ is the contribution to the
energy from the unstable modes with negative squared frequencies, and
${N}^{ad}_k(t)$ is the adiabatic particle number given by eq.(\ref{adpart}). We
have written explicitly an upper momentum cutoff, so as to absorb the
divergences in the mass and coupling renormalizations. Using the large $ N $
renormalization condition\cite{big,FRW}
\begin{equation}
{m^2_0}+ \frac{\lambda_0}{2}\phi^2 + \frac{\lambda_0}{2N} \langle
\vec{\Psi}^2\rangle = {m^2}+ \frac{\lambda}{2}\phi^2 +
\frac{\lambda}{2N} \langle \vec{\Psi}^2\rangle_R \label{rencondlargeN}
\end{equation}
and the renormalization prescription used for the effective
potential given by eqns. (\ref{massrenor}, \ref{coupren}) we find the 
once subtracted (by a time and field independent term $\propto
\Lambda^4$) renormalized energy density divided by N to be given in
the $\Lambda \rightarrow \infty$ limit by 

\begin{eqnarray}
\varepsilon & = &   
\frac{\dot{\phi}^2}{2}+\frac{\phi^2}{2} M^2(t)-\frac{M^2(t) |m|^2}{\lambda}-
\frac{M^4(t)}{2\lambda} + \varepsilon_F(t) - 
   \frac{k_u}{16\pi^2}\left(k^2_u +  M^2(t)\right)^{\frac{3}{2}}
\nonumber \\ 
&+&\frac{k_u M^2(t)}{32\pi^2}\left(k^2_u + M^2(t) \right)^{\frac{1}{2}}+
 \frac{M^4(t)}{32\pi^2}\ln\frac{\left(k_u+\sqrt{k^2_u+
M^2(t)}\right)^2}{|m|^2\sqrt{e}}  \label{eneren} \\
M^2(t) & = & -|m|^2 +\frac{\lambda}{2} \phi^2(t) + \frac{\lambda}{2N} 
\langle \vec{\Psi}^2(\vec x,t) \rangle_R \label{massagain}
\end{eqnarray}
\noindent where $M^2(t)$ is the effective time dependent mass of the
transverse fluctuations and 
\begin{equation}
\varepsilon_F(t)= \frac{1}{4\pi^2}\left\{ \int^{k_u}_0 k^2\,dk\,
\left[|\dot{\varphi}_k(t)|^2+\omega^2_k(t) \;
|\varphi_k(t)|^2 \right]+ \int^{\infty}_{k_u} k^2 \, dk \, \omega_k(t)\,
N^{ad}_k(t) \right\} \label{partprod} 
\end{equation}
contains {\em all} of the dynamical information on the spinodal
instabilities as well as of adiabatic particle production in the
broken symmetry case. This is because  the mode functions for the 
unstable modes will grow exponentially\cite{boyvega,boylee} during
the early stages of the dynamics.  The total energy is {\em finite,
conserved and real}.  

We can now make contact with the static effective potential formalism. Recall that
the static effective potential is 
a quantity defined for {\em time independent} expectation value of the field.
For this purpose, we identify the saddle point solution of the Lagrange
multiplier $M^2$ with the effective mass of the transverse fluctuations, which
in the dynamical case is $M^2(t)$ given by eq. (\ref{massagain}). Comparing the
effective potential eq. (\ref{leadordveff}) with the above expression for the
total energy, we see that we obtain the effective potential from the energy by:
i) setting $\dot{\phi}=0$ , ii) neglecting the contribution $\varepsilon_F$,
iii) setting $k_u =0$. Clearly setting $k_u=0$ leads to an imaginary part of
the effective potential whenever the effective mass is negative. This is the
region of spinodal instabilities in the case with the initial expectation value
is such that $ \eta^2_0 <2 $, and whose dynamics is described precisely by
$\varepsilon_F$.  Even in the case in which the initial condition is such that
$ \eta^2_0 > 2 $, for which $ k_u=0 $, the effective potential
completely misses the 
term $\varepsilon_F$ which is now completely determined by the adiabatic
particle production.  This is the regime of the dynamics in which particle
production is a result of parametric amplification rather than spinodal
instabilities\cite{dis,big,late}.
 
We thus see that particle production via spinodal instabilities for $ \eta^2_0
<2 $, which is accounted for in the contribution $\varepsilon_F$, solves the
problem of the imaginary part and is responsible for a {\em dynamical} Maxwell
construction. For $ \eta^2_0 > 2 $ in which case the initial state has
an energy 
larger than the top of the potential barrier and the dynamics can probe
ergodically the minima, parametric amplification leads to profuse particle
production which is accounted for by the contribution $\varepsilon_F$ now with
$k_u=0$. Therefore in both cases when particle production is important, either
in the spinodally unstable region or for large energy densities when particle
production is via parametric amplification, the effective potential provides a
misleading picture. However these are the important regions for the study of
phase transitions or large energy densities.

\subsection{ $  {\cal V}(\eta_0) >  {\cal V}(0) ,\; (  |\eta_0| > \sqrt2 )$:
symmetric evolution}

We now focus on the case of the broken symmetry potential, but when
the initial expectation value is higher in the 
potential hill than the maximum, i.e.  $  |\eta_0| > \sqrt2 $.
In this case the total energy
{\em density} is non-perturbatively 
larger than $  {\cal V}(0) $.
 Classically the expectation value of the scalar field will undergo
large amplitude 
oscillations  between the two classical turning 
points thus probing both broken symmetry states ergodically. The
classical evolution is completely {\em symmetric} in the 
sense that averaging the evolution of the expectation value over time
scales much larger than the oscillation period will 
yield a vanishing value, in contrast with the classical evolution for
$ |\eta_0|<\sqrt{2} $.  
This is not the situation
that is envisaged in   usual symmetry breaking scenarios. 
For broken symmetry situations there are no finite energy field
configurations that can sample both vacua. In the case under
  consideration with the zero mode of the scalar field with 
large amplitude and with an energy density  larger than the top of
the potential hill, there 
is enough energy in the system to sample both vacua. 

Parametric amplification transfers energy from the expectation value
to the quantum fluctuations\cite{dis,big,late}. Let us assume that 
a large fraction of the energy in the expectation value has been
transferred to the quantum fluctuations such that 
the amplitude of the expectation value now is $|\eta(\tau)| <
\sqrt{2}$. Since the quantum fluctuations react back 
on the dynamics of the expectation value and now there is an energy
density $\propto 1/g$ stored in the quantum 
fluctuations, the quantum fluctuations act as a bath with large energy
density that forces the expectation value 
to continue to sample ergodically both symmetry breaking vacuum manifolds. 

Fig. 6 displays {\em minus} the effective mass term, $ -{\cal
M}^2(\tau) $,  as a function of 
time. Clearly the effective mass is positive, translating into the
fact that $\lambda \langle \vec{\Psi}^2 \rangle/N > |m|^2$ as a
consequence of particle production via parametric amplification.
Fig. 7 displays $ \eta(\tau) $ for $ \eta(0)=5\; ; \; g=10^{-5} $. 

This situation is {\em qualitatively}
similar to that of equilibrium finite temperature field theory at
temperatures larger than the critical in the 
sense that the mean root square fluctuation of the field is larger
than the value of the field at the minima. Although 
qualitatively similar to finite temperature in equilibrium, the
situation is {\em quantitatively} different because the 
distribution function for the quantum fluctuations is {\em not}
thermal\cite{big,late}. A more proper description would be that the
total 
system composed of the expectation value and the quantum fluctuations is
described by a microcanonical ensemble at fixed energy. However treating
the quantum fluctuations as a bath in interaction with the (one) degree
of freedom described by the expectation value, one can think of an
alternative canonical description in which the expectation value
exchanges 
energy with the bath. With $\lambda \langle \Psi^2 \rangle_R/ N >
|m|^2 $ the fluctuations of the bath are large enough to allow the
expectation value to sample both broken symmetry states with equal
probability. This alternative 
canonical interpretation does not imply that the distribution functions
are thermal since there is no restriction of maximization of any form
of entropy.   
 
The evolution of the expectation value is
damped because of the transfer of energy to all the other modes via
the mean field.  In the present
dynamical case, this  `symmetry restoration' is just a consequence 
of the fact that there is a  large energy density in the initial
state, much larger than the top of the 
tree level potential, thus under the dynamical evolution the system
samples both vacua equally.  We note however that symmetric evolution
would ensue even in the 
classical theory without quantum backreaction simply because the
energy density is large enough. The initial state is {\em not} a
broken symmetry state. It has a  large enough energy density  to allow
the expectation value to overcome the potential barrier.   

Thus the criterion for symmetric evolution   when the
tree level potential allows 
for broken symmetry states is simply that the energy density in the initial
state be larger than the top of the 
tree level potential. That is when the initial amplitude of the
expectation value is 
such that $  {\cal V}(\eta_0) >  {\cal V}(0) $.  
In this case the dynamics is similar to the unbroken
symmetry case, the amplitude of the 
zero mode  damps out, transferring energy to the quantum
fluctuations via parametric amplification and non-linear resonances\cite{late}.
We see that the amplitude  of the zero mode tends asymptotically
to zero very slowly in a manner similar to the unbroken symmetry
case ($ m^2 > 0 $)\cite{big,late} while it 
oscillates symmetrically around  $\eta=0$.

These points are clearly illustrated  in figs. 6 and 7,  $ {\cal
M}^2(\tau)  $ and   $ \eta(\tau) $ for $ 
\eta_0 =   5 > \sqrt2 $ [and hence $   {\cal V}(\eta_0) >  {\cal V}(0) $], 
where $  {\cal V}(\eta) $ is given by eq.(\ref{defVeta}) 
and $ g = 10^{-5} $. We find the typical behaviour
of unbroken symmetry\cite{big,late}. It is clear from the figure the
presence of two frequencies: one corresponds to twice the asymptotic
meson mass $ {\cal M}(\infty) $.

We see from fig. 6 that $ {\cal M}^2(\tau)  $ tends asymptotically to 
$$
{\cal M}^2(\infty) = -1 + \frac12 \, \eta_0^2 \; .
$$
This value follows from  the unbroken symmetry result \cite{late} upon
changing the sign of the tree level mass squared $ +1 \to -1 $. 

We emphasize again, that the fact that we find symmetric evolution for
a negative 
tree level mass squared is not a surprising result, it is due to the
high energy of the initial states 
considered $ \sim {\cal O}(1/g) >> 1 $. Since the dynamical evolution
sampled both vacua symmetrically  from the 
beginning (even in the classical theory),  there never was  symmetry
breaking in 
the first place, and `symmetry restoration'   is just the statement
 that the initial state has enough 
energy density so that the {\em dynamics}  probes  both vacua symmetrically
 despite the fact that the tree level potential allows for broken
symmetry ground states. 

Energy conservation and mode mixing through the mean field leads to a damped
evolution for the expectation value but non-perturbative particle
production via parametric amplification, results in a highly 
excited quantum state in which the energy has been distributed mainly
throughout the states in the parametrically unstable
band(s)\cite{big,late}. The sign of the tree level mass squared which
is so crucial near the ground state becomes here almost irrelevant. 

\section{Quantum Phase Ordering Dynamics}
Phase ordering refers to the process phase separation after a sudden
cooling from the disordered, high temperature phase into the low temperature
phase. During this process the system breaks up in ordered domains, this
is the essence of Kibble's mechanism\cite{kibble,kibble2,vilen} for
the formation of defects after a phase transition. Whereas the dynamics of phase
ordering is fairly well understood in condensed matter
physics\cite{bray,marco}, a microscopic quantum field theoretical
treatment is still missing. Analogies with condensed matter physics had been used
to provide some dynamical estimates in quantum field theory\cite{zurek},
but a full non-equilibrium, {\em dynamical} treatment beginning from
a microscopic quantum field theory is just beginning to
emerge\cite{boylee,rivers,beilok}. Analogies with the condensed matter
situation led to the study of {\em classical} scaling solutions for the dynamics
of texture collapse and relaxation in the non-linear sigma model in a
matter-radiation dominated cosmology\cite{turok,filipe}.  

In this section we provide a description of the process of phase
ordering, implementing the non-equilibrium formulation in the full
quantum field theory (QFT). 

Consider the situation in which the  
system has been quenched from a symmetric (disordered), high
temperature phase into the low temperature (ordered) 
broken symmetry phase. 

In QFT this quench can be introduced by
considering an initial quantum mechanical wave function(al) prepared
at $t=0$  centered on top of the potential, i.e. a wave
function(al) that is localized 
near the origin in field space\cite{erickwu,boylee,boyvega}. This wave
functional is then evolved in time with the Hamiltonian with a potential of the
broken symmetry form. This wave function(al) will evolve in time
by spreading out to sample the equilibrium states, which results in a
growth of correlations and fluctuations. It is precisely in this
situation that the density matrix formulation and its probabilistic
interpretation offer the most enlightening description.  
The early time dynamics relaxing the quench assumption has been been
recently studied in reference\cite{bowick}. 

 In the large $ N $ limit the concept of topological defect is rather
ambiguous, and as we will see, the important concept is the  formation
of long-wavelength, large amplitude (semiclassical) field
configurations during the evolution.  

Let us consider the situation of a critical quench, with initial
conditions of vanishing expectation value and 
first derivative at the initial time, i.e. $\eta(\tau=0)= 0 \; ; \;
\dot{\eta}(\tau=0)=0$. Since this is a fixed point 
of the dynamics for the expectation value, it will remain zero all
throughout the evolution. 
From the probability distribution given by eq.(\ref{probability}) and
the initial conditions on the mode functions
(\ref{unsfrequ})-(\ref{stafrequ}),  
we see that at the initial time the most probable configurations are
those whose spatial Fourier transform are of the 
order $1/\sqrt{\Omega_q}$. Therefore the probability for finding
semiclassical long wavelength field configurations with
non-perturbatively large amplitudes, i.e. $\Psi_k \propto 1/\sqrt{g}$,
in the quantum  ensemble is exponentially suppressed by the inverse of
the coupling constant, i.e. ${\cal P}[\Phi_{sc}]\propto e^{-1/g}$.

However, from the
evolution equation for the mode functions (\ref{modokR}), we see that
the modes with wavevectors in the spinodally unstable 
band grow exponentially and their amplitude becomes of order
$1/\sqrt{g}$ at the spinodal time $\tau_s \sim \ln(1/g)$. Figures 
(11)-(12) show $ \sqrt{g} |U_q(\tau=100,200)| $ vs. $ q $ respectively for
$ g=10^{-7}\; ; \; \eta(\tau=0)=\dot{\eta}(\tau=0)=0 $. 
We infer that for $\ tau > \tau_s $ there is a {\em unsuppressed}
probability for finding long wavelength $ k\leq |m| \; ( q<1 ) $ 
large amplitude $ \propto 1/\sqrt{g} $ field configurations in the
ensemble. The interpretation of this observation is 
that spinodal instabilities lead to the formation of {\em
non-perturbative semiclassical long-wavelength field configurations
with amplitude} $\sim 1/\sqrt{g}$ in some functional direction. This
is consistent with the fluctuations sampling the minima of the 
potential, that is $ \langle \vec{\Psi}^2\rangle/N \sim |m|^2/\lambda $
or mean square root fluctuation of the field $ \sim
|m|/\sqrt{\lambda} $.  

An important concept in the dynamics of phase ordering is that of the
time dependent correlation length. In condensed 
matter systems\cite{bray} the process of phase ordering at late times
(after initial transients shortly after the 
quench) is described in terms of correlation functions that show a
{\em scaling} form in terms of a dynamical correlation 
length which at long times is independent of the (short) initial
correlation length in the disordered phase. A scaling solution has
also been found\cite{turok} in the {\em classical} large $ N $ limit but
in matter or 
radiation Friedman-Robertson-Walker cosmology. 

We now turn our attention to a detailed study of the equal time correlation
function to understand the emergence of scaling.  

\subsection{Correlation Functions}

Consider the case of a critical quench 
i.e. $\langle \vec{\Phi} \rangle = \langle \vec{\dot{\Phi}} \rangle=0$
, i.e. $\eta(0)=\dot{\eta}(0)=0$. In terms of the 
dimensionless coordinate distance $\vec{r} = \overline|m| \vec{x}$
and the dimensionless quantities introduced before 
(\ref{overlinem}-\ref{gren}) the equal time correlation function becomes

\begin{eqnarray}\label{fcorr}
< \Phi^a({\vec x}, t) \, \Phi^b(\vec 0,t) > &=& C( r,\tau)\;
\delta^{a,b} \cr \cr
&=& \overline{|m|}^2 \; \delta^{a,b} \int { d^3q \over {(2\pi)^3}} \; 
e^{i {\vec q}. {\vec r}} \; |U_q(\tau)|^2 
 = { 1 \over { 2\pi^2  r }}\int_0^{\infty} q \, dq \, \sin(q  r) \;
|U_q(\tau)|^2 \; .
\end{eqnarray}
\noindent with $ r= |\vec r| $.

We display in fig. 10, $ g r C(r,\tau) $ as a function of $ r $ for
various values of $ \tau $. This figure reveals several remarkable features 
that we analyze in detail below. 

\subsection{Intermediate  time: $\tau \leq \tau_s$}

For times $ \tau $ earlier than the nonlinear or spinodal time $
\tau_s $ we can 
neglect the back-reaction of the quantum fluctuations in the evolution
equations for 
the mode functions. This is the linear regime and we can use the
spinodal behavior  
of the mode functions in order to describe the behavior of $ C(r,\tau)
$. When $\tau 
\approx  \tau_s $, the quantum fluctuations  become non-perturbatively
large and the 
back-reaction, encoded in the term $g\Sigma(\tau)$ in the evolution equations
becomes comparable to the tree level term.  For weak coupling, the
spinodal time scale is given by\cite{late}

\begin{equation}
\tau_s \approx \frac12 \log\left[ \sqrt{8 \over {\pi}}\frac1{g}\right]
+ \frac34  \log \log\left[ {8 \over {\sqrt{\pi}\;g}}\right] + \ldots
\; . \label{spinotime}
\end{equation}
and the mode functions for $ \tau < \tau_s $ can be reliably approximated as
\cite{late}
$$
U_q(\tau) = {1 \over {2 \sqrt{1 - q^4}}}\left\{ \left[ 
\sqrt{1 - q^2}-i (1+q^2) \right]\; e^{\tau \sqrt{1 - q^2}} + 
{\cal O}\left(e^{-\tau \sqrt{1 - q^2}}\right) \right\}\; .
$$
for wavevectors in the spinodally unstable band $ 0 < q^2 < 1 $. 
Inserting this expression into eq.(\ref{fcorr})
yields upon saddle point integration\cite{boylee},
\begin{equation}\label{ctempr}
C(r,\tau) = { 1 \over { 16 \, (\pi \; \tau)^{3/2} }} \; e^{2\tau - {r^2
\over { 4 \tau}}} \left[ 1 + {\cal O}\left({r^2 \over \tau^2}\right)
\right]\; . 
\end{equation}
Eq.(\ref{ctempr}) is in remarkable agreement with our
numerical calculations in its range of validity\cite{late}. 
The exponential decay with  $ r $ in eq.(\ref{ctempr}) can be
interpreted as the emergence  of a {\em dynamical} correlation length
\begin{equation}
\xi_e = 2 \sqrt{\tau} \label{qftcorr}
\end{equation}
The correlation function in this regime can be written in the form
\begin{equation}
C(r,\tau) = g(\tau) F\left[\frac{r}{\xi_e(\tau)}\right]. \label{scaling}
\end{equation}
which, however, is not a true scaling form, since the function $g(\tau)$ is not a power of the dynamical correlation length. 
This behavior is reminiscent of a diffusive process and
similar to the results of phase ordering kinetics for 
non-conserved order parameter in the non-relativistic case\cite{bray},
which will be briefly reviewed in the next section. The
`diffusive' type behavior in this case is understood from the fact
that the modes that grow the 
fastest are those for which $q<<1$ for which a non-relativistic
approximation of the modes is valid.  

As we will see below, however, the
$\sqrt{\tau}$ behavior of the dynamical correlation length  
is only valid for $\tau \leq \tau_s$; there is a crossover to a
linear scaling law and a true scaling solution at longer times.

\subsection{Short Distance Analysis}

The correlator for small distances can be obtained by expanding 
$\sin(q r)$ in eq.(\ref{fcorr}) and restricting the integration to
the band of spinodally unstable momenta 
$q\leq 1$, since the contributions
from $ q > 1 $ are perturbatively small corrections.
Doing this yields:
$$
C(r,\tau)\buildrel{r \to 0}\over= { 1 \over { 4\pi^2  }}\left[
\Sigma(\tau) - { r^2 \over 6} \int_0^1 q^4 \, dq \;
|U_q(\tau)|^2 +  {\cal O}\left(r^4\right) \right]\; .
$$

For late times we can use the sum rules arising from energy conservation\cite{late}
$$
g \Sigma(\infty) = 1 \quad , \quad \int_0^1 q^4 \, dq \;
|U_q(\infty)|^2= \frac1{4g} \; .
$$
Therefore,
$$
C(r,\tau)\buildrel{r \to 0, \tau>> 1}\over= { 1 \over { 4\pi^2 \, g }}\left[
1 - \frac{1}{24} \, r^2 +  {\cal O}\left(r^4\right) \right]\; .
$$
This behaviour is in perfect agreement with our numerical results for short
distances.

\subsection{Large distance and late time behavior}
For $r >>1$ the integral is dominated by the small q-region, and since
the only important modes are those in 
the spinodally unstable band $q \leq 1$, changing variables to $ Q=qr $
the correlation function can be expressed as
\begin{equation}
rC(r,\tau) = \frac{1}{2\pi^2 r^2}\int^{\infty}_0 Q \, dQ \; \sin
Q\; |U_{\frac{Q}{r}}(\tau)|^2 \label{largert} 
\end{equation}

For $ r>>1 $, only the mode functions $ U_q(\tau) $ with very small $
q $ are relevant. Their long time behavior can be obtained from the
following reasoning. At late times $ \tau >> \tau_s $ the effective
mass term vanishes asymptotically as given by eq. (\ref{latemass})\cite{late}. 

Since the effective mass vanishes asymptotically at long time as given
by (\ref{latemass})\cite{late}  there are no secular terms in the
perturbative solution for the $ q=0 $ mode\cite{late} so that the
mode with $ q=0 $ must behave as\cite{late} 
\begin{equation}
U_0(\tau) = L +K\tau +\mbox{small oscillatory terms} \label{zeromodelate}
\end{equation}
\noindent The Wronskian guarantees that neither of the complex
coefficients, $L,\ K$ can vanish\cite{late}. 
Figs.(11)-(12) show both the real ($ \sqrt{g} \; U_{R,0}(\tau)$) and
imaginary ($\sqrt{g} \; U_{I,0}(\tau)$) parts of $\sqrt{g} \; U_0(\tau)$
respectively for $\eta(0)=\dot{\eta}(0)=0$; 
$ g=10^{-7} $ where it can be seen that $ L \sim K \sim {\cal
O}(1/\sqrt{g}) $. This linear growth with time signals
the onset of a novel form of Bose Einstein condensation, in the sense
that at very large times the zero momentum mode will become macroscopically
occupied\cite{late}. Since the  mass term vanishes asymptotically, the
solutions to the equations of motion will be of the form 
$$
U_q(\tau) \buildrel{\tau \to \infty}\over= A_q \; e^{iq\tau}+ B_q \;
e^{-iq\tau} 
$$
The equation of motion for the modes is analytic in the momentum
variable $ q $, therefore analyticity in $ q $ fixes the small $q$
behavior at long times to be of the form 
\begin{equation}
U_{q << 1}(\tau) = L\,\cos(q\tau)+\frac{K}{q}\,\sin(q\tau).
\label{smallqmodes} 
\end{equation} 
Figs. (8)-(9) show the behavior of the mode functions at two
large times, for very small $ q $, the behavior of the modes as a
function of $ q $ is very well described by the form (\ref{smallqmodes})
with a numerical error 
in the small $ q $ region less than a few percent.  

At large time and distance ($ r;\tau >>1 $) the correlation function
(\ref{largert}) becomes 

\begin{equation}
r\; C(r,\tau) \buildrel{r>>1, \tau \to \infty}\over= \frac{|K|^2}{2\pi^2
}\int^{\infty}_0 \frac{dQ}{Q} 
\sin Q \sin^2\left(Q\; \frac{\tau}{r}\right)= \frac{|K|^2}{8\pi
}\;\Theta(2\tau -r) \; .\label{largert2} 
\end{equation}
The correlation function falls off as $ 1/r $ for $ r<
2\tau $ and vanishes for $ r>2\tau $. which is precisely 
the behavior found numerically and shown in detail in fig. 10 which
reveals that the correlation function vanishes for 
\begin{equation}\label{causalidad}
r> 2(\tau-\tau_s)+c  
\end{equation}
with $ \tau_s $ the spinodal time and $ c \approx 2$ for small coupling.  
A similar analysis using the mode functions (\ref{smallqmodes}) leads to
the $ \cos(2\tau)/\tau $ behavior of $ {\cal M}^2(\tau) $ at large times.

The $1\slash r$ fall-off can be interpreted as a correlation function of {\em
massless} Goldstone modes. However, these are {\em not} free modes, as their
correlator would have a $1\slash r^2$ fall off. 

The interpretation is the following, for times $ \tau > \tau_s $ there
is a zero momentum condensate formed by Goldstone bosons travelling at
the speed of light and  back-to-back. That is, massless particles
emitted from the points   $  (0,\tau) $ and $ (r,\tau) $
form  propagating fronts which at time $ \tau $ are at a distance $ \tau -
\tau_s $ from the origin and from  $ r $, respectively. 

These space-time points are causally connected for
$  2(\tau-\tau_s) \geq r $. Otherwise, the correlator
vanishes. Causality allows the addition of a positive constant $ c $ 
as in eq.(\ref{causalidad}).

The root mean square fluctuation of the field 
inside this propagating front is non-perturbatively large $ \sim
|m|/\sqrt{g} $ reflecting the process of phase ordering. The fact that
the root mean square fluctuation of the field inside these domains
means that although 
the expectation value of the order parameter vanishes, inside these growing
domains the fluctuations sample the broken symmetry minima. These then are
ordered domains with an average expectation value inside each domain to
be given the mean square root fluctuation and therefore saturated at the
equilibrium values at the minima of the potential. 

For $ r< 2(\tau-\tau_s)+c $ we find numerically that
\begin{equation}
C(r,\tau) \approx \frac{0.1}{g \; r}
\end{equation}
\noindent for $ \eta(0)=\dot{\eta}(0)=0 $ and small coupling. 
This result is in complete agreement with the estimates 
(\ref{largert2}) since from figures (11)-(12) (linear growth) we infer
that $ g|K|^2 \approx 2 $. 

Inserting the low $ q $ behaviour of the mode functions
(\ref{smallqmodes}) into eq.(\ref{sigmafin}) for yields for late times
$$
 \Sigma(\tau) \buildrel{\tau \to \infty}\over=  \Sigma(\infty)
+ {\cal O}\left( {\sin 2\tau \over \tau}
\right)\; .
$$
Comparing this with the results of ref.\cite{late}, we see that this
approximation yields the correct asymptotic behaviour for $
\Sigma(\tau) $ except for the logarithmic phases in the $ 1/\tau $
term [see eq.(\ref{latemass})]. 

The asymptotic form of the equal time two-point correlation function given
by eqn. (\ref{largert2}) reveals a {\em true} scaling solution.
Introducing  the dynamical correlation length $\xi(\tau) \approx \tau$ and
defining the variable $x=r/\xi(\tau)$ 
 it can be written in the form
\begin{equation}
C(r,\tau) = \left[\xi(\tau)\right]^{-2(1-z)} F[x]
\end{equation}
with the anomalous dynamical exponent $z=1/2$ (the naive scaling length dimension of the field is 1) and the scaling function 
\begin{equation}
F[x]\propto \frac{1}{x} \Theta(2-x)
\end{equation} 

This scaling solution reveals the non-perturbative buildup of a {\em large
anomalous dimension}. 

\subsection{Correlation length}

We can compute the correlation length at time $\tau $ using the
standard expression, 

$$
\xi(\tau) \equiv {\int d^3r \; |r| \; C(r,\tau)\over \int d^3r \;  C(r,\tau)}
$$

There are two different time regimes: i) the early-intermediate time
$\tau \leq \tau_s$ and ii) the late time $\tau >> \tau_s$ with
different behavior for the correlation length. Thus there is a
crossover in the scaling behavior of 
the correlation function at the spinodal time scale $\tau_s$, with the
correlation length behaving as 
\begin{equation}
\xi(\tau) \propto \left\{ \begin{array}{cc}
\sqrt{\tau} & \mbox{for $\tau \leq \tau_s$} \\
\tau-\tau_s &  \mbox{for $\tau \geq \tau_s$}
\end{array} 
\right. \label{crossover}
\end{equation}

For times larger than the spinodal time, when the spinodal
fluctuations have been shut-off by the growth of the self-consistent
field, and the dynamics is driven by the non-linear evolution, 
the correlation length grows monotonically with time
becoming infinite as $t\rightarrow \infty$; this is a consequence of
the presence of 
massless particles. Moreover, the vanishing of the correlation
function for $ r>2 \tau $ is consistent with causality and tells us
that the correlation length must be of 
the order of magnitude given by eq.(\ref{crossover}) for at long times. 

Only for  infinite time the correlation length takes the value infinite
as one would expect in the presence of massless particles. The time effectively
acts as a infrared cutoff making $ \xi $ finite for finite time.

\subsection{Where are the Defects?}

As mentioned above, there are no defects as such in the $O(N)$ model
in the large 
$N$ limit. However, what we are seeing is that this is not necessarily
the relevant 
question, in terms of the phase ordering. What we should look for are long
wavelength, {\em non-perturbatively} large configurations\cite{losa}.

These are always present in the ensemble described by the density matrix of the
system. However, if our system were just a thermal one, such configurations would be
highly suppressed, since their energies would be proportional to $1\slash \lambda$.

What we see in our evolution, though, is that spinodal instabilities drive the
growth of long wavelength, non-perturbatively large configurations. We should then
expect to find them with unsuppressed probability in the ensemble after the spinodal
time, say. A way to quantify this is to use the probability distribution obtained
from the density matrix.

The probability of finding a particular field configuration with 
spatial Fourier transform $\Psi_k$ at time $t$ in the ensemble is
given by

\begin{equation}
{\cal P}[\Psi,t] = \prod_{\vec{k}}
\frac{{\cal{N}}_{\Psi,k}(t_0)}{\sqrt{2}|\varphi_k(t)|} 
\exp\left\{- \frac{\vec{\Psi}_k \cdot
\vec{\Psi}_{-k}}{2|\varphi_k(t)|^2}\right\} \label{probability} 
\end{equation}
which shows clearly that at a given time $t$, the most likely configurations to
be found in the ensemble are those whose spatial Fourier transform are given by
$\varphi_k(t)$. 

The mode functions $\varphi_k(t)$ become non-perturbatively large, 
$\propto 1/\sqrt{\lambda}$ because of spinodal unstabilities. Therefore even
when at the initial time field configurations with non-perturbatively large
amplitudes are exponentially suppressed for weak couplings, at times of $t_s
\propto \ln(1/\lambda)$ large amplitude field configurations will be
unsuppressed in the quantum ensemble.

This interpretation is further supported by the results of reference\cite{mottola}. 

\subsection{A classical but {\em stochastic} description}
At the end of the stage of linear instabilities at $\tau \approx \tau_s$
we have seen that the mode functions in the spinodally unstable band
$0<q<1$ attain non-perturbative large amplitudes $\sim 1/\sqrt{g}$, whereas
the larger wavelength modes remain perturbatively small. All of the
correlation functions are dominated by the modes that achieved
non-perturbatively large amplitudes through spinodal
decomposition. This suggests that the field should be decomposed as the sum of a {\em
classical} part corresponding to the amplified modes and a quantum
mechanical part corresponding to those modes with perturbative
amplitudes 
\begin{eqnarray}
\vec{\Psi}(\vec x,t) & = & \vec{\Psi}_{sc}(\vec x,t)+
 \hat{\vec{\Psi}}(\vec x,t) \nonumber \\ 
 \vec{\Psi}_{sc}(\vec x,t) & = & 
\int_{\vec k,|k|<|m|}\frac{d^3k}{(2\pi)^{3/2}}\;
\left[ {\vec {A}}_{\vec k}\; {\varphi}_{\vec k,sc}(t)\; e^{i
 \vec k \cdot \vec x}  + {\vec {A}}^{*}_{\vec k}\; {\varphi}^*_{\vec
 k,sc}(t)\; e^{-i \vec k  \cdot \vec x}  \right] 
\label{semiclass} \\
\hat{\vec{\Psi}}(\vec x,t) & = & 
\int \frac{d^3k}{(2\pi)^{3/2}}\;
\left[ \hat{\vec {a}}_{\vec k}\;  {\varphi}_{\vec k,qm}(t)\;  e^{i \vec k
 \cdot \vec x}  + \hat{\vec {a}}^{*}_{\vec k}\;  {\varphi}^*_{\vec
 k,qm}(t)\;  e^{-i \vec k  \cdot \vec x}  \right] \nonumber
\end{eqnarray} 
\noindent where $ {\varphi}_{\vec k,sc} $ are the 
$ {\cal O}(1/\sqrt{g}) $ 
parts of the mode functions, whereas $ {\varphi}_{\vec k,qm} $ are the
$ {\cal O}(1) $, i.e. perturbatively small part. Clearly the split
between the semiclassical and quantum components of 
the field is ambiguous. However,  this ambiguity is of $ {\cal O}(g) $
corresponding to extracting the small components of the mode functions
and therefore remains perturbatively small.  

Whereas the $ {\hat {\vec a}} $ are quantum mechanical operators with usual
commutation relations, the variables $ \vec{A} $ are classical. However
in order to reproduce the non-perturbative contributions of the
correlation functions these variables 
must be described in terms of a {\em Gaussian stochastic probability
distribution}  so that  
\begin{eqnarray}
&& <<{\vec {A}}_{\vec k}>> = <<{\vec {A}}^*_{\vec k}>> =0 \nonumber \\
&& <<{\vec {A}}_{\vec k}{\vec {A}}^*_{\vec k'}>> = \delta_{\vec k,\vec
k'} \label{stochastic}  
\end{eqnarray}
The stochasticity of these semiclassical variables is necessary to
reproduce all of the correlation functions that are obtained from
averages with the 
{\em quantum mechanical} density matrix given by
(\ref{psidensitymatrix}). Because in the 
large $ N $ limit the quantum density matrix is Gaussian, the
semiclassical probability distribution function is also Gaussian.  

We see that a stochastic description arises naturally from the time-evolution of the
system, and is {\em not} put in by hand. 

\subsection{Summary of quantum phase ordering dynamics}
We summarize here the results from this
section to provide a picture of the process of quantum phase ordering in the
large $ N $ limit for the case of a critical quench,
i.e. $\eta(0)=\dot{\eta}(0)=0$.  

\begin{itemize}
\item{The process begins with the exponential growth of spinodally unstable modes. The equal space-time  two point correlation function grows and 
begins to compete with the tree level terms in the equations of motion. This
is the region of linear instabilities that lasts up to the spinodal
time $t_s \approx 1/|m| \; \ln(1/g)$ when $g\Sigma(\tau) \approx 1$. The correlation function is of the scaling form, with a time dependent correlation length that scales
as $\xi(t) \propto \sqrt{t}$. The growth of quantum fluctuations are 
interpreted as the formation of semiclassical coherent condensates with
wavelengths $k \leq |m|$, with unsuppressed probability of being represented
in the quantum ensemble. } 

\item{ At $t \sim t_s$ the quantum backreaction of the modes produces 
a crossover and for $t> t_s$ the evolution is
non-linear\cite{late}. The effective 
mass of the fluctuations vanishes and the distribution functions reveal
 non-perturbative long-wavelength condensates of Goldstone bosons. A novel
form of Bose Einstein condensation is manifest in the linear growth
with time of the zero momentum mode function. This soft condensate of
Goldstone bosons in turn leads to a correlation function that falls
off as $1/r$ and is cut-off by causality  at $r > 2(t-t_s)$. We
interpret this result as that the 
process of phase ordering occurs via the formation of a domain or bubble that
grows at the speed of light for $t> t_s$ inside which there is a
non-perturbative, semiclassical condensate of Goldstone bosons. Inside
these 
domains the mean square root fluctuations are $\sim
|m|/\sqrt{\lambda}$ and therefore sampling the equilibrium minima of
the potential, i.e. the order 
parameter averaged over this domain saturates at the equilibrium value.
The equal time correlation function falls off as $1/r$ in contrast to
free field behavior $ 1/r^2 $.}

\item{ For $t \geq t_s$ the non-perturbative aspects of the dynamics
can be described by a semiclassical but stochastic distribution function.
The semiclassical description is achieved through  the field expansion in terms
of the modes in the spinodally unstable band with coefficients that are
interpreted as stochastic variables with a Gaussian probability
distribution function. This description reproduces the
non-perturbative contributions 
to all of the correlation functions in the large $ N $ limit.}  

\end{itemize}

\section{Comparison with classical Phase Ordering Kinetics:}
We now compare the results obtained above on quantum phase ordering
dynamics to those previously found in the literature but in {\em
different contexts} 
but within {\em classical stochastic} field theory: i)
Non-relativistic condensed matter: phase ordering kinetics for
non-conserved order parameter in the large $ N $
limit\cite{bray,marco}, ii) Cosmology: the relaxation of 
texture-like configurations in a Friedmann-Robertson-Walker background
metric in the non-linear sigma model in the large $ N $ limit\cite{turok}. 
Our main purpose is to compare and contrast our results to those
obtained within the {\em classical} large $ N $  approximation and to point out
that the body of results found here describe a fundamentally different 
set of phenomena and scaling. 

\subsection{Phase ordering kinetics in condensed matter:}
The process of ordering after a quench in condensed matter physics has
been studied both theoretically and experimentally very intensely during
the last decade (for a thorough review see\cite{bray}). 
Amongst the theoretical approaches, the large $ N $ limit stands out as one
of the few that yields to an exactly solvable description of the
dynamics and used as a yardstick against which to compare other 
approaches\cite{bray}. Let us now briefly summarize the most
relevant features of the solution in the large $ N $ limit. For non-conserved 
order parameter, the process of phase ordering is studied by means of
the time dependent  Ginsburg-Landau  equation (TDGL)

\begin{equation}
\frac{\partial \vec{\phi}(\vec x,t)}{\partial t}=
-\Gamma \frac{\delta F}{\delta \vec{\phi}(\vec x,t)}. \label{tdgl}
\end{equation}
Here $\Gamma$ a phenomenological dissipative coefficient, which in this 
model can be absorbed in a definition of time, and
the Ginzburg-Landau free energy $F$ is given by
\begin{equation}
F= \int d^3x \left[ \frac{1}{2}\left(\vec{\nabla}\vec{\phi}(\vec x,t)
\right)^2 + V(\vec{\phi}(\vec x,t))\right] \label{freenergy}
\end{equation}
with $V(\vec{\phi})$ given by eq.(\ref{barepotential})(with
$m^2_0 <0$ for broken symmetry).
The TDGL equation is {\em purely dissipative} and a simple calculation of
the energy, which is given by total spatial integral of the free
energy density, shows that it is a monotonically decreasing function of time. 

In terms of the rescaled field $\vec{\phi}^2=(2N|m_0|^2/\lambda_0)\,
\vec{\eta}^2$ and rescaled time and spatial variables $|m_0|^2 \Gamma
t = \tau \; \; ; \; \;  |m_0|\vec x=\vec z $ the TDGL equation now reads

\begin{equation}
\frac{\partial \vec{\eta}}{\partial \tau} = \nabla^2 \vec{\eta}+
\vec{\eta}-(\vec{\eta})^2 \vec{\eta} \label{dimlesstdgl}
\end{equation}
The large $ N $ limit is implemented by a Hartree-like  factorization of the
non-linear term\cite{bray} $\vec{\eta}^2 \rightarrow \langle
\vec{\eta}^2 \rangle$ where the average is over the initial
probability distribution function. The initial state is assumed to be
completely disordered and described by a Gaussian distribution function with
\begin{equation}
\langle \vec{\eta}(\vec z,0) \rangle=0 \; \; ; \; \; \langle
\eta^a(\vec z,0)\; \eta^b(\vec z',0) \rangle  = {\Delta}\; \delta^{a,b}  \; 
\delta^3(\vec z - \vec z') \label{inicorreta} 
\end{equation}
\noindent with $\Delta$ a constant determined by the high temperature
short ranged correlations\cite{bray,marco}. This factorization leads
to a simple equation 
\begin{eqnarray}
\frac{\partial \vec{\eta}}{\partial \tau} & = &  \nabla^2 \vec{\eta}+
{\cal{M}}^2(\tau)\;  \vec{\eta} \nonumber \\
{\cal{M}}^2(\tau) & = & 1- \langle \vec{\eta}^2 \rangle \label{effmass}
\end{eqnarray}
but with a self-consistency condition much in the same
manner as the large $ N $ description in quantum field theory, notice
that the effective mass term ${\cal{M}}^2(\tau)$ is similar to that
found in eq. (\ref{masef}). However, notice that eqs.(\ref{effmass})
are not Lorentz invariant but only Galilean invariant.

The large $ N $ solution of the TDGL equation leads to the following
asymptotic expressions for the effective mass and correlation
function\cite{bray}  

\begin{eqnarray}
&& {\cal{M}}^2(\tau) \buildrel{\tau \to \infty}\over=
\frac{3}{4\, \tau} \label{masstdgl} \\
&& \langle \eta^a({\vec z},\tau)\;
 \eta^b(\vec 0, \tau) \rangle = \delta^{a,b} \; S_o \;
 \exp\left(-\frac{|\vec z|^2}{8 \tau}\right) \label{corrtdgl}
\end{eqnarray}
\noindent where again the averages are over the initial probability 
distribution and $S_o$ depends on the initial conditions. 

Furthermore, recently a detailed study of the asymptotic behavior in
time in the large $ N $ limit of phase ordering kinetics in these
models revealed a form of Bose Einstein condensation of long
wavelength fluctuations asymptotically in time\cite{marco}.  

The TDGL description leads
asymptotically to Goldstone 
bosons, with an effective mass term that vanishes asymptotically as in
eq.(\ref{masstdgl}), whereas in QFT the effective mass vanishes as in
eq. (\ref{latemass}). The oscillatory behavior of (\ref{latemass}) is
an important difference because it is responsible for non-perturbative
anomalous dimensions\cite{late}. The dynamical correlation length
obtained from the TDGL equal time correlation function,
eq. (\ref{corrtdgl})  agrees  
(up to an overall scale) with that obtained in  QFT for $\tau <
\tau_s$ and given by eq.(\ref{qftcorr}). However, whereas the TDGL
correlation length 
holds at all times (after some short initial transients\cite{bray,marco}),
the QFT solution exhibits a crossover at $\tau > \tau_s$ [see
eq. (\ref{crossover})] to a linear dependence determined 
by causality and the onset of a Bose-Einstein condensate. This
crucially different behavior is related to the non-relativistic
nature of the TDGL equations in contrast with the fully relativistic
invariance of our model. 

A remarkable similarity is that both the QFT version and the TDGL one
of the large $ N $ model dsiplay the onset of a novel form of a Bose
Einstein condensate. In QFT we have seen that the
$ q=0 $ quantum mode grows linearly with time as a 
consequence of a vanishing effective mass term. This condensation mechanism
is responsible for cutting off the spatial correlations for $ r>2t $ as
discussed above, a result compatible with causality. In the TDGL
description, a Bose condensate is found\cite{marco} in order to
satisfy 
the consistency condition in the asymptotic equilibrium state, this
condensate results in a Bragg peak at zero spatial
momentum\cite{marco}. 
An important {\em difference} is that whereas the TDGL description is
purely dissipative and the asymptotic dynamics is rather insensitive to
the initial conditions and coupling, in the QFT description
dissipation is compatible with energy conservation and time reversal
invariance and the asymptotic evolution depends on the  initial
conditions and coupling.  

\subsection{Textures and scaling in a FRW background:}
In references\cite{turok} the process of phase ordering in radiation
and matter dominated  FRW cosmological backgrounds has been studied at the
classical level in the non-linear sigma model (NLSM). A more detailed
connection between the cosmological 
setting and the TDGL description described above has been provided in
reference\cite{filipe} and we refer the reader to those references for
more details. We here highlight the most important ingredients of the
solution to compare to our results on quantum phase ordering kinetics which
obviously are valid for Minkowski space-time. The main evolution equation
studied in refs.\cite{turok,filipe} are given by (using the
same notation for the variables)
\begin{equation}
\frac{\partial^2 \vec{\eta}}{\partial \tau^2}+
\frac{\alpha(\tau)}{\tau} \frac{\partial \vec{\eta}}{\partial
\tau}-\nabla^2 \vec{\eta}+ {\cal M}^2(\tau)\; \vec{\eta}=0 \label{nlsmeqn}
\end{equation}
\noindent where $\tau$ now refers to {\em conformal} time, $\alpha(\tau)=
2 \; d\ln a(\tau)/d\ln \tau $ and $ a(\tau) $ the FRW scale factor as a
function of conformal time. Here the effective mass ${\cal M}^2(\tau)$
arises from 
a spatial Gaussian average of the non-linear term in the equation of 
motion of the non-linear sigma model\cite{turok,filipe}. The scaling
solution found in references\cite{turok,filipe} result in the
self-consistent behavior for the effective mass term given by 
\begin{equation}
{\cal M}^2(\tau) = \frac{{\cal M}^2_0}{\tau^2} \label{nlsmmass}
\end{equation}  
which leads to an exact form for the solution of the mode functions (in
terms of comoving wavevectors) in terms of Bessel
functions\cite{turok,filipe}. The solutions are of the scaling form 
\begin{equation}
\vec{\eta}_{\vec k}(\tau) \propto F(|\vec k|\tau)\label{scalingnlsm}
\end{equation}

As shown  in\cite{filipe} these gaussian scaling solutions lead to
equal (conformal) time correlation functions that are cutoff at $r = 2\tau$
and that reveal a dynamical (comoving) correlation length that grows linearly
in conformal time $\xi \sim \tau$. 
In radiation and matter dominated FRW cosmologies the size of the
causal horizon $\sim a(\tau)\;\tau $ in conformal time 
and the cutoff in the correlation function is a consequence of causality. 

The solution of refs.\cite{turok,filipe} have no $ \alpha = 0 $
limit. The normalization and the power spectrum have an infrared
divergence for $ \alpha \rightarrow 0 $ which makes impossible to
fulfil the NLSM constraint in that limit. Thus, our quantum Minkowski
solution is not related to the classical gaussian solutions in FRW for
the NLSM. Furthermore, our treatment is fully quantum whereas a
classical stochastic approach is used in refs.\cite{turok,filipe}. 

There are only two aspects where our quantum results and the solution
of refs.\cite{turok,filipe} are similar: a correlation length that grows
linearly with time and a correlation function with support  inside $
r<2\tau $. 

We find a crossover from the spinodal regime to the non-linear 
regime. Of course this crossover cannot be captured by the NLSM
since in the NLSM the field is constrained to lie at the minima of
the potential, i.e. there is no possibility of spinodal instabilities or
rolling of the expectation value. In addition, our  effective mass
squared decreases as $ 1/\tau $ while it oscillates with frequency $
2 \, m $ whereas in the NLSM the effective squared mass vanishes as $
1/\tau^2 $.  This different behaviour corresponds to the different
nature of the IR divergences in FRM and Minkowski space-times.

\section{Conclusions}
In this article we have obtained a novel set of non-perturbative and
non-equilibrium phenomena in the leading order in the large N limit of
a scalar theory. We have elucidated the shortcomings of a static effective
potential description in the large N limit and contrasted
this formulation to the results obtained within a consistent and
non-perturbative treatment of the dynamics out of equilibrium in the
same approximation. Whereas the 
static effective potential is complex in the region of coexistence, the
non-equilibrium evolution leads to a {\em dynamical Maxwell construction}:
all the expectation values in the coexistence region are allowed asymptotic
solutions of the dynamics and the ``restoring force'' for the evolution
of the expectation value vanishes asymptotically, i.e. a flat {\em dynamical} potential in the coexistence region. The dynamical Maxwell construction is a consequence of non-perturbative particle production through spinodal
instabilities. The particles are produced mainly at low momentum. A dynamical correlation length $\xi(t)$ emerges whose behavior as a function of time reveals a crossover from $\xi(t) \propto \sqrt{t}$ for
$t< t_s \approx |m|^{-1}\ln(1/\lambda)$ to $\xi(t) \propto (t-t_s)$
for $t> t_s$. For $t>t_s$ the evolution is dominated by the non-linearities
and a true scaling solution emerges for which the field scales with a
non-perturbatively large anomalous dynamical exponent $z=1/2$. These
novel phenomena in the non-linear regime are a consequence of the onset
of a non-equilibrium Bose-Einstein condensate at zero momentum that 
grows linearly in time. This condensate and the density of modes nearby
ensure that the equal time two-point correlation function vanishes by
causality for $r > 2(t-t_s)$. We have studied the process of {\em quantum}
phase ordering and interpreted the long-time phase ordering dynamics as
proceeding via the growth of  domains or bubbles that expand at the speed
of light inside which there is a non-perturbative condensate of Goldstone
bosons. The equal time two-point correlation function falls off as $1/r$ inside these
domains. 

 The analysis in terms of a quantum density matrix leads to a  description of the
non-equilibrium ensembles and correlations in terms of a classical but stochastic distribution function which describes large amplitude configurations with unsuppressed probabilities. We have also compared our
results of quantum phase ordering dynamics to those obtained in {\em stochastic} field theories both in condensed matter
(non-conserving TDGL equation) as well as the non-linear sigma model in a Friedmann-Robertson-Walker cosmology establishing
some qualitative similarities and important differences.   

Furthemore we have also clarified aspects of the dynamics when the initial energy density is {\em larger} than that of the top of  the potential. Our analysis shows how the expectation value is
driven to zero, showing unambiguously that the dynamics is O(N) symmetric as a consequence of non-perturbative particle production via parametric amplification.

\section{Acknowledgements:} 
D.B and H.J. de V. thank J. Alexandre, E. Branchina,  C. Destri, Y. Kluger,
E. Mottola, J. Polonyi, R. Rivers and N. Tetradis for stimulating discussions,
D.B. thanks M. Zannetti and 
M. Hindmarsh for  useful conversations. D. B. thanks the N.S.F for
partial support through grant award: PHY-9605186  and LPTHE for warm
hospitality.  R. H., is supported by DOE grant DE-FG02-91-ER40682. We
thank NATO for partial support.




\begin{figure}
\centerline{ \epsfig{file=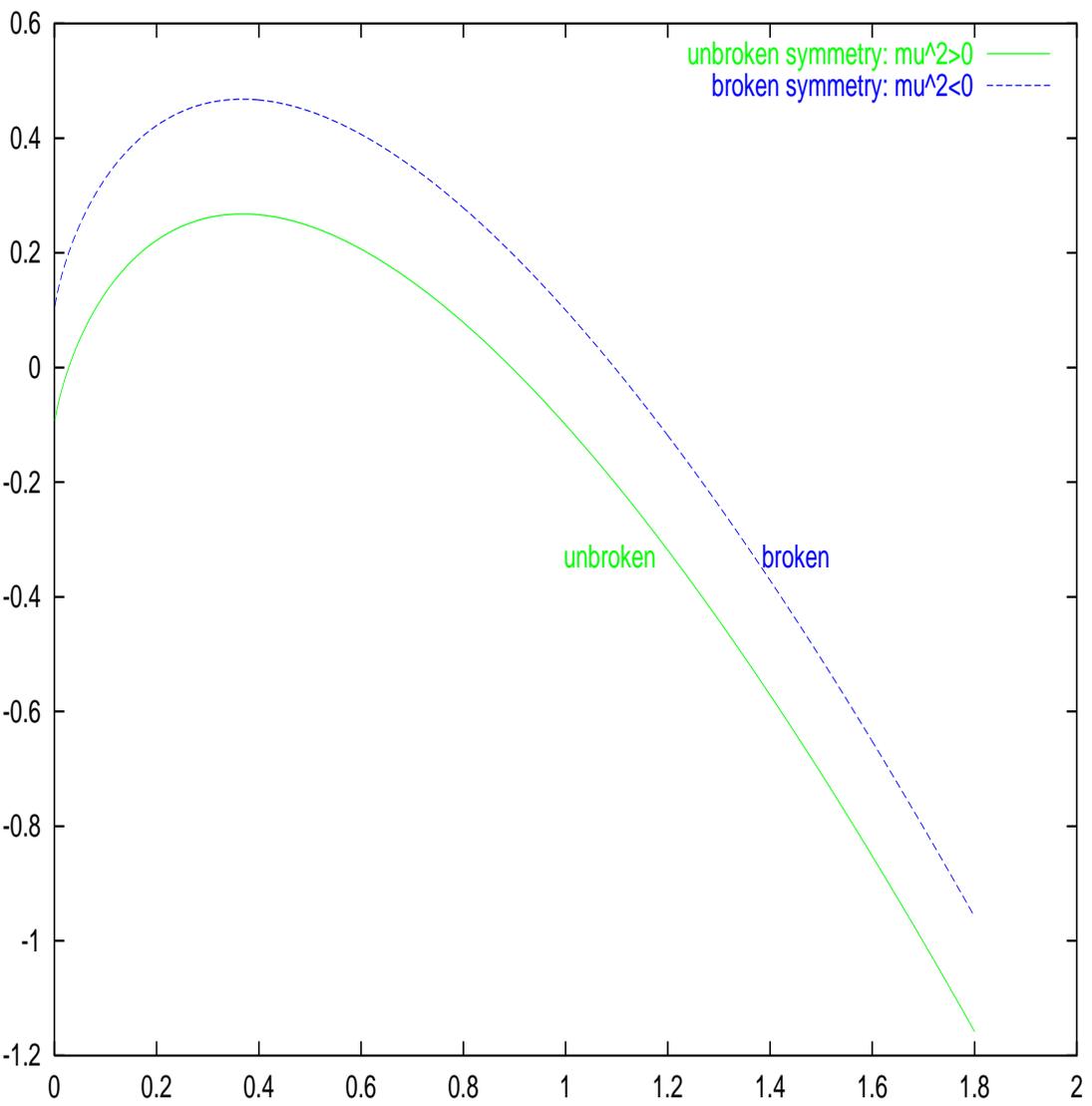,width=6in,height=6in}}
\caption{The dimensionless scalar field $ \varphi^2(w) $ as a function
of the dimensionless mass $ w $ for a coupling $ \lambda$ such that 
$ {32 \pi^2  \over \lambda} \; e^{-32\pi^2/\lambda} = 0.1 $. See
eq.(\ref{fiw}). 
\label{fig2}}
\end{figure}


%
\begin{figure}
\centerline{ \epsfig{file=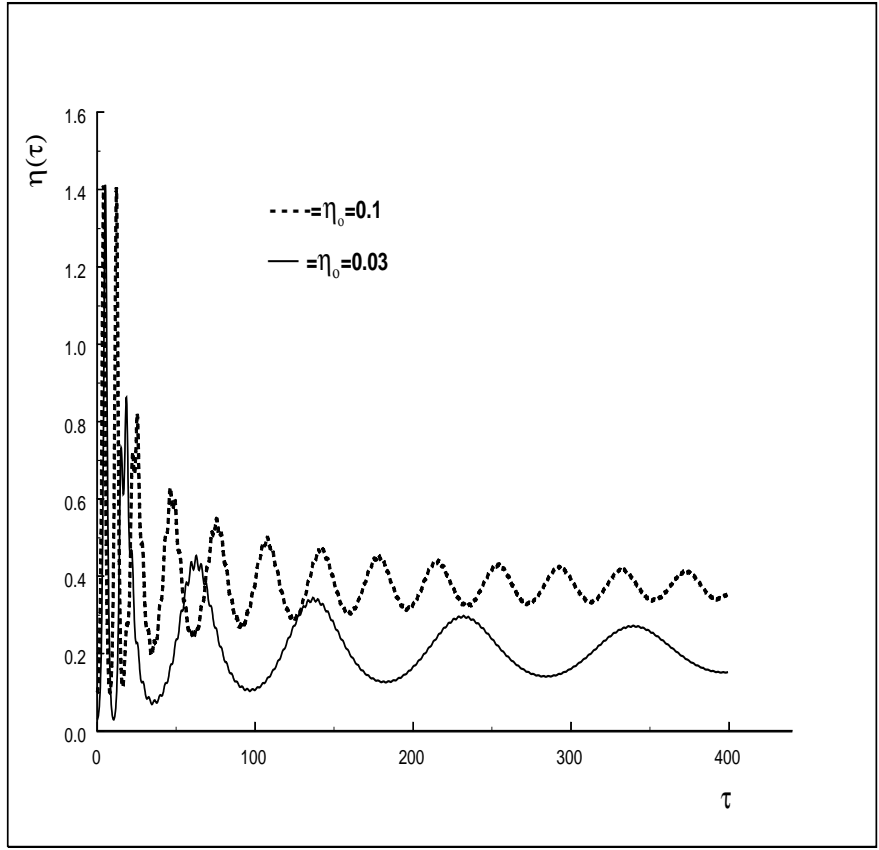,width=5in,height=5in}}
\caption{$\eta(\tau)$ vs. $\tau$. Solid line corresponds to $\eta_0=0.03\; ; \; \dot{\eta}_0=0$,
dashed line corresponds to $\eta_0=0.1\; ; \; \dot{\eta}_0=0$, for $g=10^{-7}$. \label{fig3}}
\end{figure}
%


\begin{figure}
\centerline{ \epsfig{file=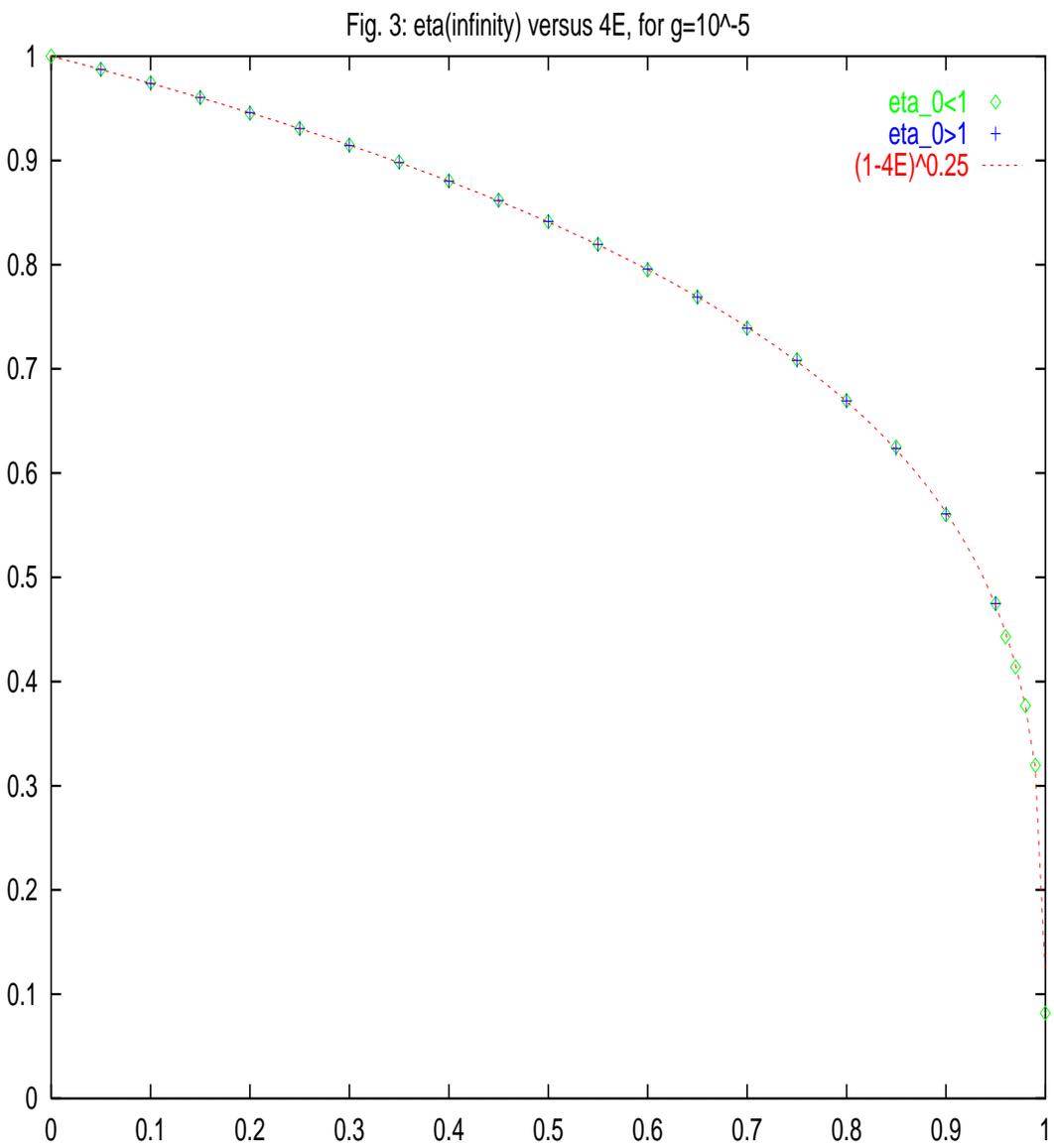,width=6in,height=6in}}
\caption{$ \eta_{\infty} $ as a function of $ 4 E = (\eta_0^2 - 1)^2 $. The
points $ + $ correspond to $ \eta_0 > 1 $ whereas the points $ \Diamond $
correspond to $ \eta_0 < 1 $. As discussed in section IV, $
\eta_{\infty} $ is invariant under the exchange $ \eta_0 \Leftrightarrow
\sqrt{2 - \eta_0^2} $. The continuos curve is $
\eta_{\infty} = \left[  \eta_0^2 \, (2 -  \eta_0^2)
\right]^x $ with  $ x = 0.25 $. 
\label{fig4}}
\end{figure}


\begin{figure}
\centerline{ \epsfig{file=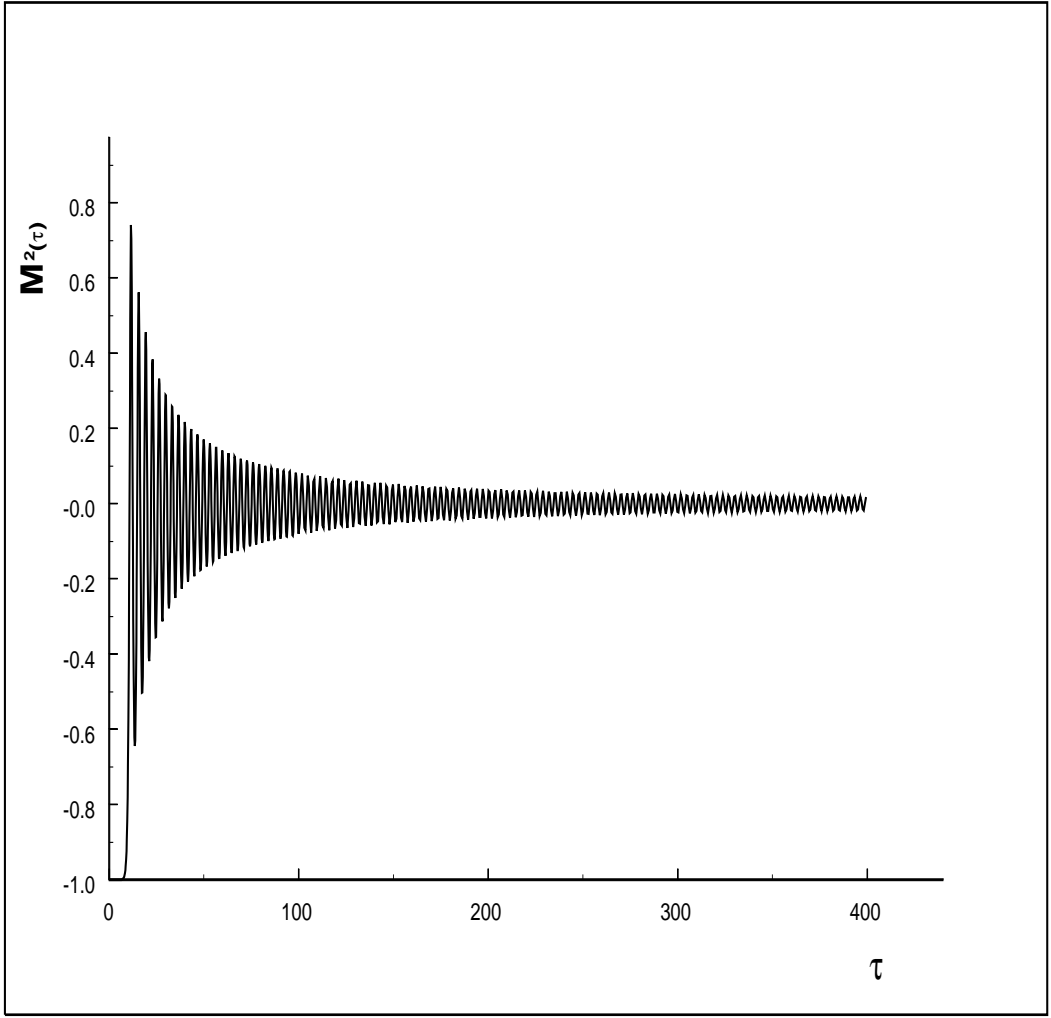,width=6in,height=6in}}
\caption{${\cal M}^2(\tau)$ vs. $\tau$ [see eq.(\ref{masef})] for
$ \eta(0)=\dot{\eta}(0)=0$, $g=10^{-7} $ \label{fig5}}
\end{figure}


\begin{figure}
\centerline{ \epsfig{file=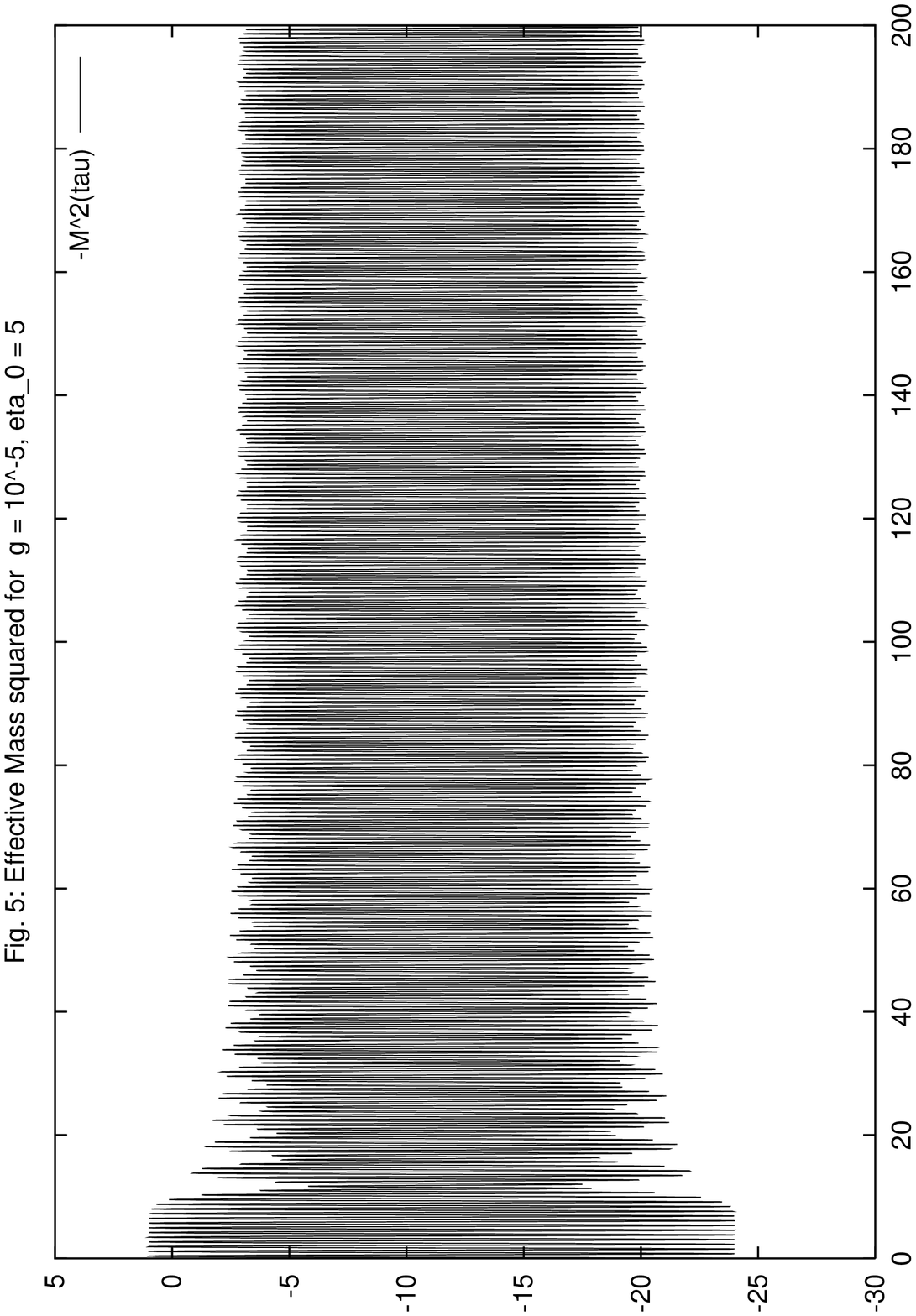,width=6in,height=6in}}
\caption{$ {\cal M}^2(\tau) $ as a function of $\tau$ for $ g =
10^{-5} $ and $\eta_0 = 5 $.  
\label{fig6}}
\end{figure}


\begin{figure}
\centerline{ \epsfig{file=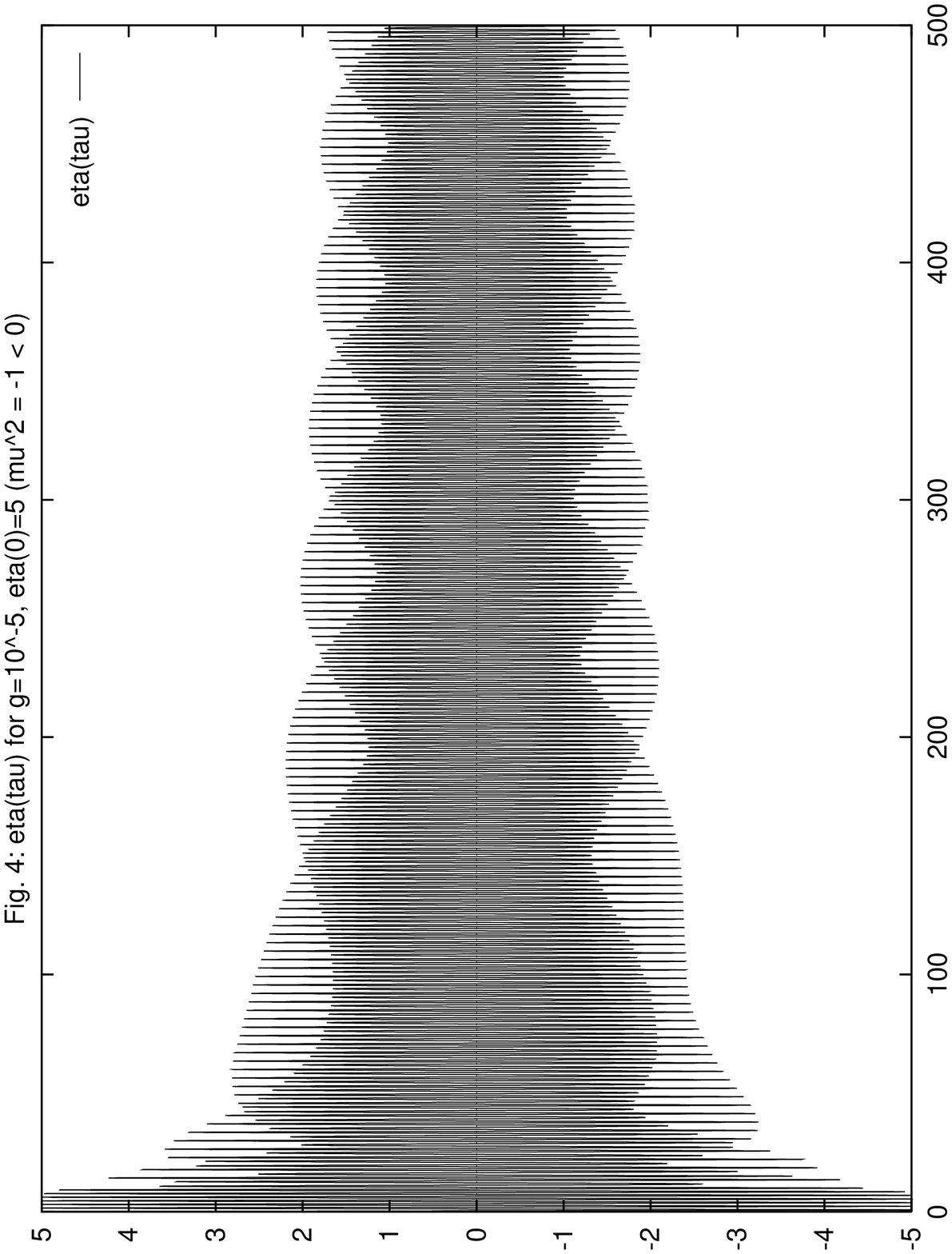,width=6in,height=6in}}
\caption{$ \eta(\tau) $ as a function of $\tau$ for $ g = 10^{-5} $ and $
\eta_0 = 5 $. 
\label{fig7}}
\end{figure}


\begin{figure}
\centerline{ \epsfig{file=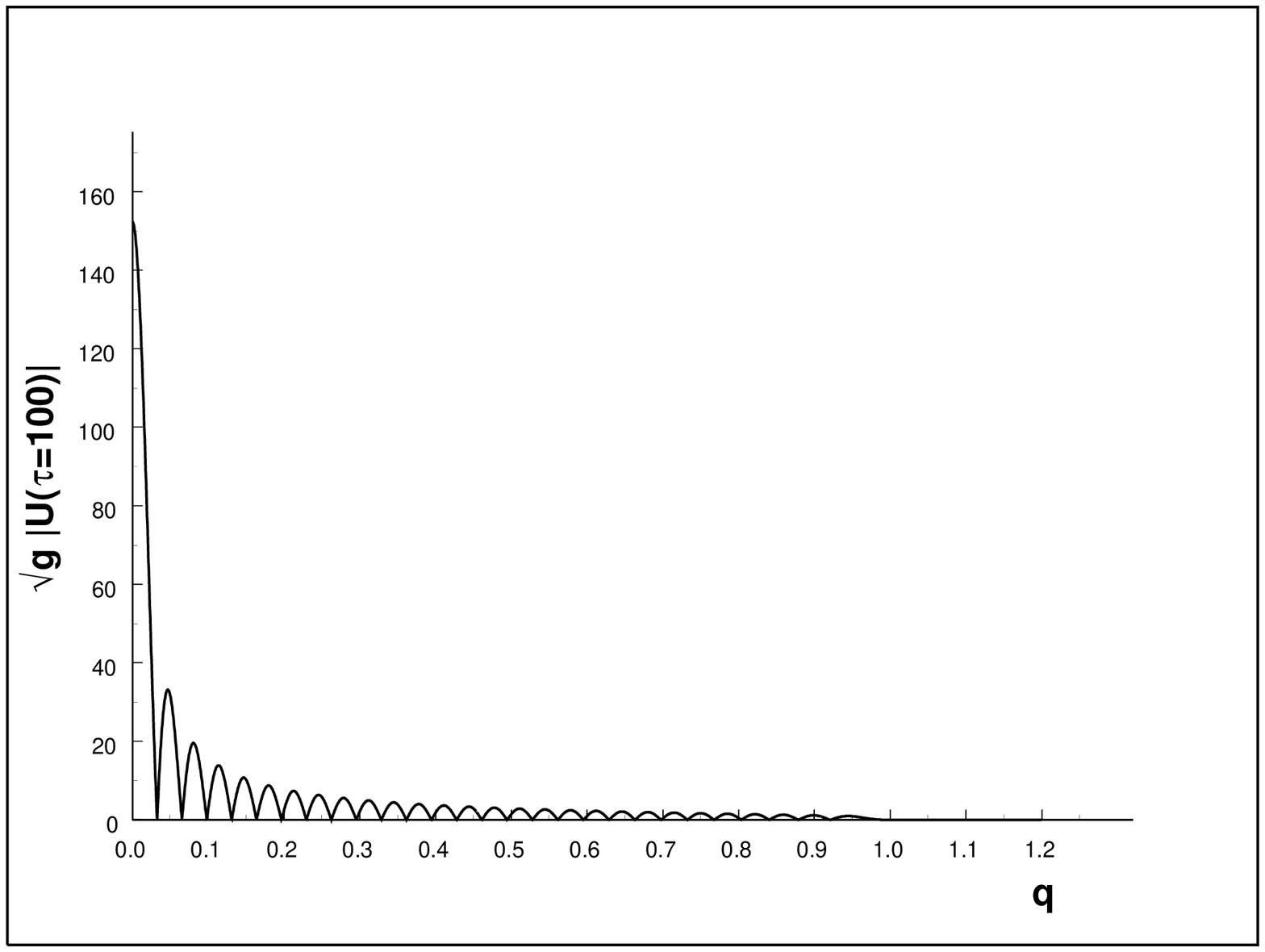,width=6in,height=6in}}
\caption{$\sqrt{g}|U_q(\tau=100)|$ vs. $q$ (see eq.(\ref{modokR}) for
$\eta(0)=\dot{\eta}(0)=0$, $g=10^{-7}$ \label{fig8a}}
\end{figure}


\begin{figure}
\centerline{ \epsfig{file=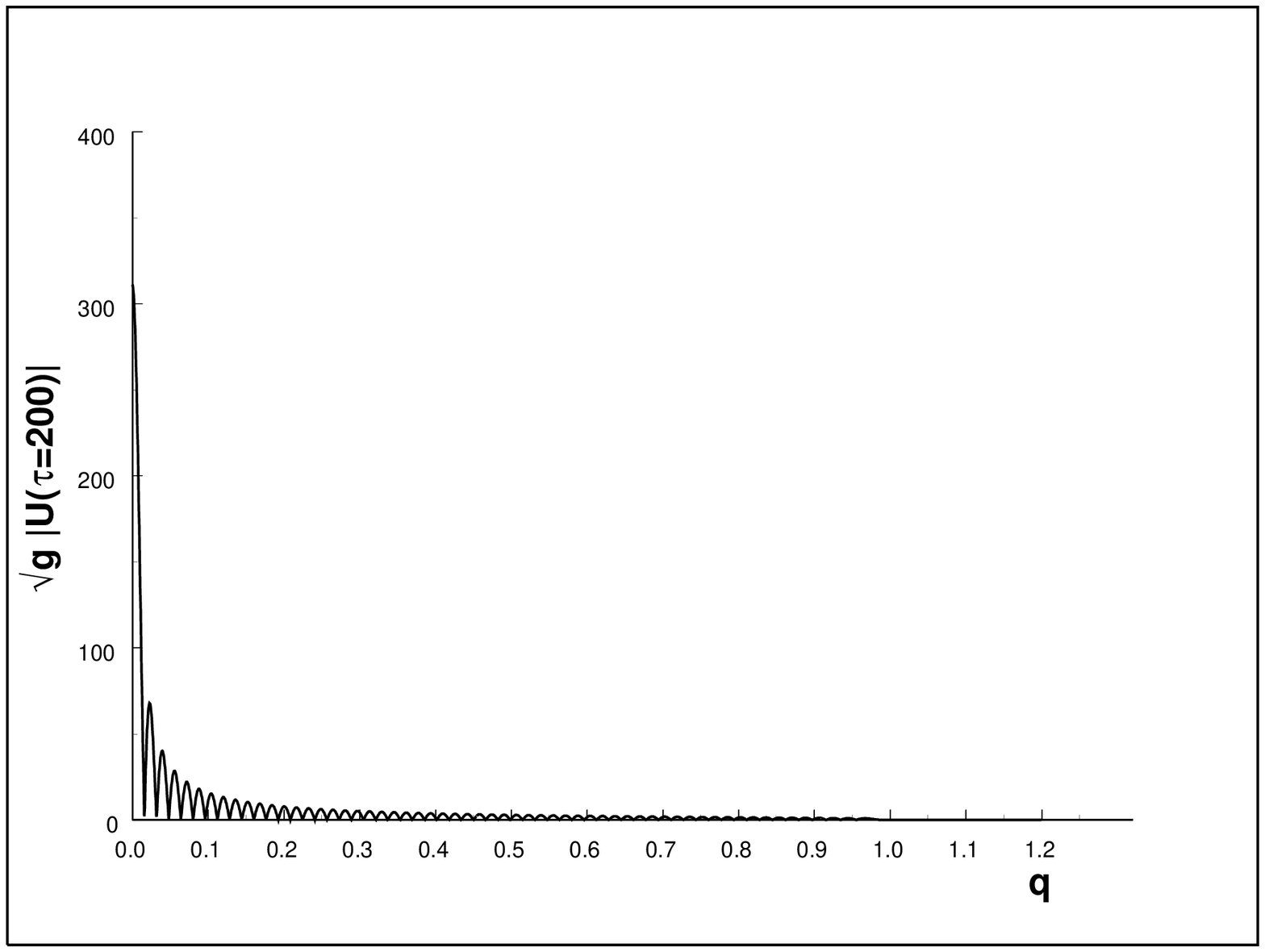,width=6in,height=6in}}
\caption{$\sqrt{g}|U_q(\tau=200)|$ vs. $q$ (see eq.(\ref{modokR}) for
$\eta(0)=\dot{\eta}(0)=0$, $g=10^{-7}$ \label{fig9}}
\end{figure}



\begin{figure}
\centerline{ \epsfig{file=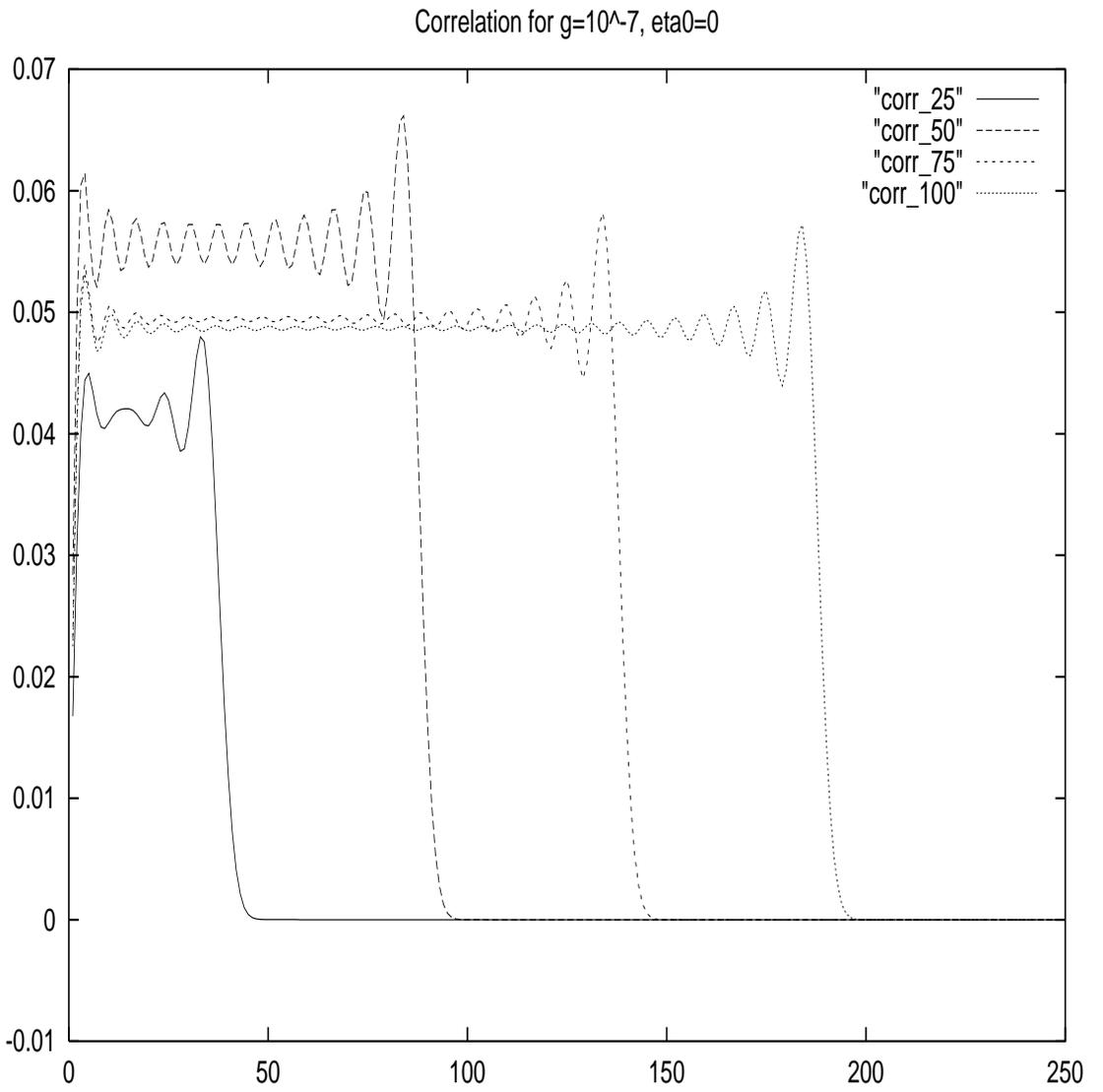,width=6in,height=6in}}
\caption{$ g r\, C(r,\tau) $ as a function of $ r $ for $ \tau = 25 \;
; 50 \; 75 ; \; 
100  $ for $ g = 10^{-7} $ and $ \eta_0 = \dot{\eta}(0)=0 $.
 \label{fig10}}
\end{figure}


\begin{figure}
\centerline{ \epsfig{file=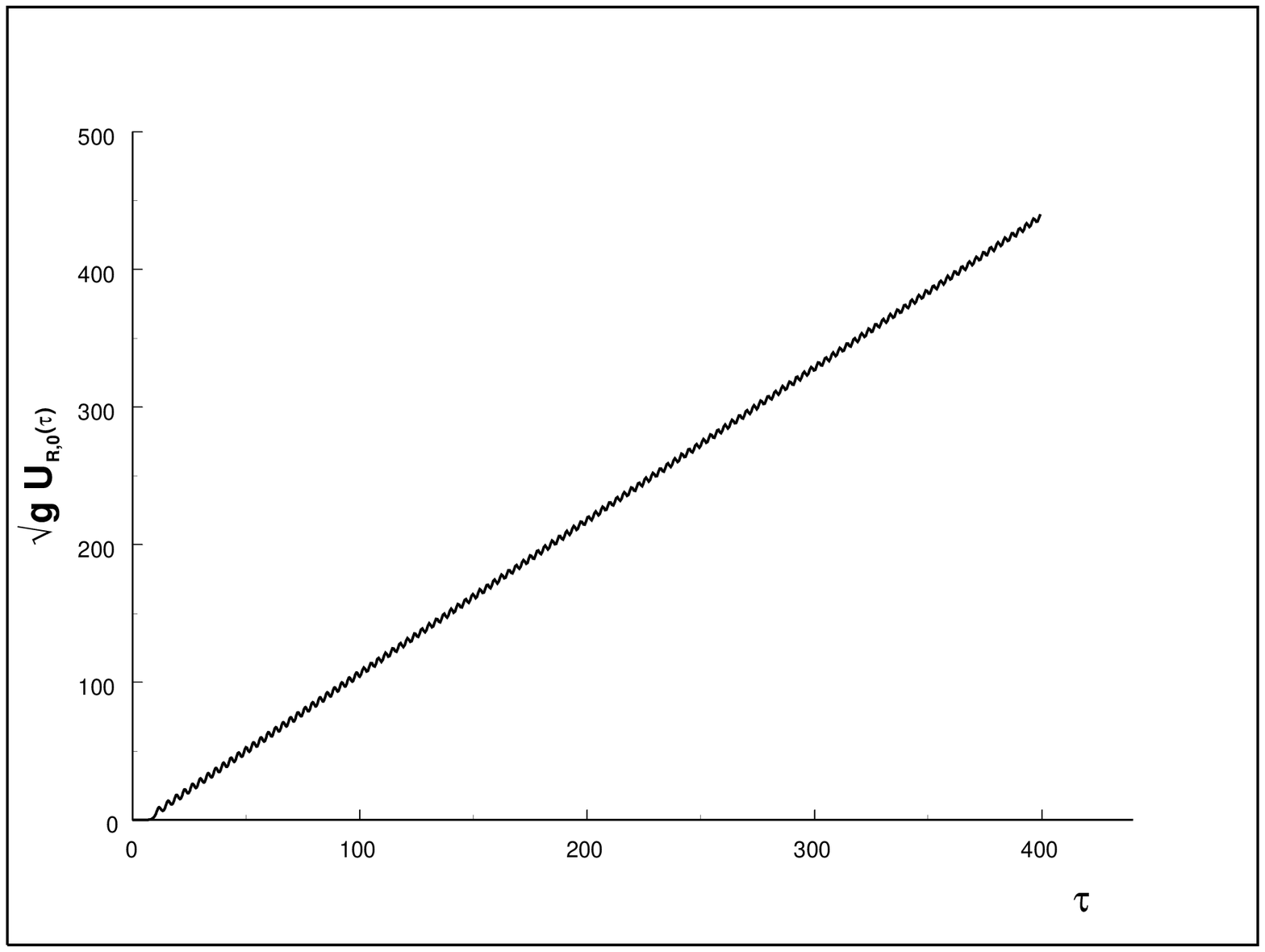,width=6in,height=6in}}
\caption{$\sqrt{g}U_{R,q=0}(\tau)$ vs. $\tau$  for
$\eta(0)=\dot{\eta}(0)=0$, $g=10^{-7}$ \label{fig11}}
\end{figure}


\begin{figure}
\centerline{ \epsfig{file=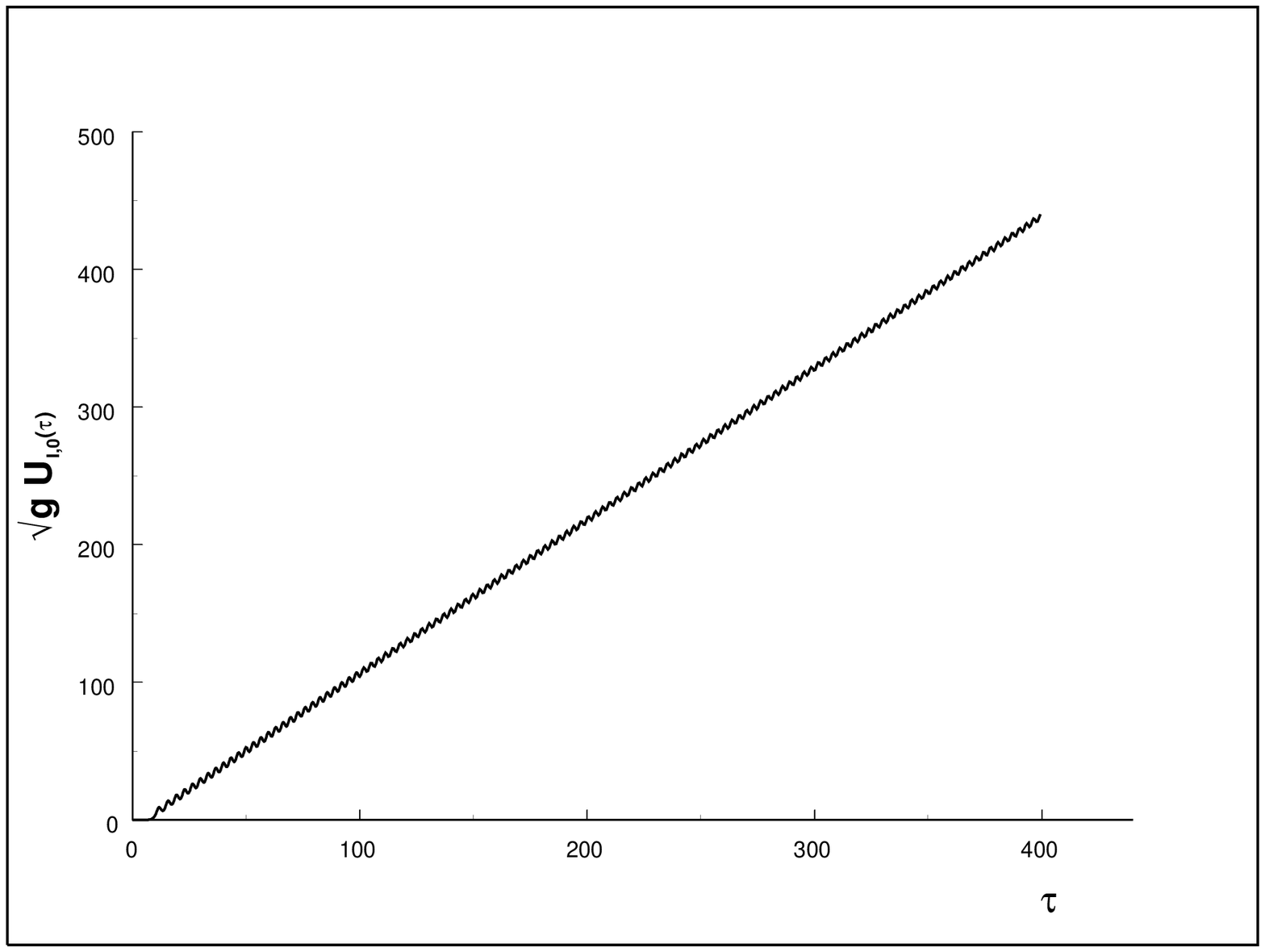,width=6in,height=6in}}
\caption{$\sqrt{g}U_{I,q=0}(\tau)$ vs. $\tau$  for
$\eta(0)=\dot{\eta}(0)=0$, $g=10^{-7}$ \label{fig12}}
\end{figure}



\begin{thebibliography}{99}
\bibitem{dis} 
D. Boyanovsky, H. J. de Vega, R. Holman, D.-S. Lee and A. Singh, 
Phys. Rev. {\bf D51}, 4419 (1995). 
D. Boyanovsky, H. J. de Vega and R. Holman, Proceedings of
the Second Paris Cosmology Colloquium, Observatoire de Paris, June 1994,
pp. 127-215, H. J. de Vega and N. S\'anchez, Editors (World
Scientific, 1995); Advances in Astrofundamental Physics, Erice
Chalonge School, N. S\'anchez and A. Zichichi Editors, (World
Scientific, 1995). 

\bibitem{linon}
D. Boyanovsky, H. J. de Vega and R. Holman, 
 Vth. Erice Chalonge School, Current Topics in Astrofundamental
 Physics, N. S\'anchez and A. Zichichi Editors, World Scientific,
 1996, p. 183-270. 
D. Boyanovsky, M. D'Attanasio,
H. J. de Vega, R. Holman and D. S. Lee, 
Phys. Rev. {\bf D52}, 6805 (1995).

\bibitem{mottola}  F. Cooper, S. Habib, Y. Kluger, E. Mottola,
 Phys. Rev. D55 (1997), 6471.


\bibitem{tsu} D. Boyanovsky,   H. J. de Vega, R. Holman, S. Prem Kumar and
 R. D. Pisarski, Phys. Rev. {\bf D57}, 3653 (1998).

\bibitem{lindekov} L. Kofman, A. Linde and A. Starobinsky, Phys. Rev. Lett. 73,
3195 (1994) and 76, 1011 (1996). A. Linde,
Lectures on Inflationary Cosmology, in
Current Topics in Astrofundamental Physics, `The Early Universe', Proceedings
of the Chalonge Erice School, N.  S\'anchez and A. Zichichi Editors, Nato ASI
series C, vol. 467, 1995, Kluwer Acad. Publ. 

\bibitem{big}
 D. Boyanovsky, H. J. de Vega, R. Holman and J. Salgado,
Phys. Rev. {\bf D54}, 7570 (1996);  D. Boyanovsky, D. Cormier,
 H. J. de Vega, R. Holman, A. Singh, M. Srednicki; Phys. Rev. D56
 (1997) 1939. 

\bibitem{kolb}  E. W. Kolb, A. Riotto and Igor I. Tkachev,
Phys. Rev. D56 (1997) 6133;  E.W. Kolb, A. Riotto and I.I. Tkachev,
Phys. Lett. B423 (1998) 348.  

\bibitem{tkachev}  S. Khlebnikov and I. Tkachev,
Phys. Rev. Lett. 77, 219 (1996) and  79, 1607 (1997);
I. Tkachev, S. Khlebnikov, L. Kofman, A. Linde, hep-ph/9805209; 
 A. Riotto and I. I. Tkachev
 Phys.Lett. B385 (1996) 57;  I. I. Tkachev, 
 S. Yu. Khlebnikov, I. I. Tkachev, Phys. Lett. B390 (1997) 80.


\bibitem{kawa} S. Kasuya and M. Kawasaki, Phys. Rev. D56, 7597 (1997)
and D58 (1998);  Phys. Lett. B388, 686 (1996); M. Yamaguchi,
J. Yokoyama and M. Kawasaki, Prog. Theor. Phys. 100 (1998) 535.

\bibitem{baacke1} J. Baacke, K. Heitmann and C. Paetzold,
Phys. Rev. D55 (1997) 7815 and D58 (1998) 125013.

\bibitem{late} D. Boyanovsky, H. J. de Vega, C. Destri, R. Holman and
J. Salgado, Phys. Rev. {\bf D57}, 7388 (1998).

\bibitem{losa} 
F. Cooper, S. Habib, Y. Kluger, E. Mottola, J. P. Paz, P. R. Anderson,
Phys. Rev. {\bf D50}, 2848 (1994). 
F. Cooper, Y. Kluger, E. Mottola, J. P. Paz, Phys. Rev. {\bf D51},
2377 (1995); F. Cooper and E. Mottola, Mod. Phys. Lett. A 2, 635 (1987);
F. Cooper and E. Mottola, Phys. Rev. D36, 3114
(1987); F. Cooper, S.-Y. Pi and P. N. Stancioff,
Phys. Rev. D34, 3831 (1986).  

\bibitem{baacke2} J. Baacke, K. Heitmann and C. Paetzold,
Phys. Rev. D57 (1998) 6406;  D57 (1998) 6398 and  D56
(1997) 6556. 

\bibitem{FRW} D. Boyanovsky, H. J. de Vega and R. Holman, 
Phys. Rev.  {\bf D 49}, 2769 (1994); 
D. Boyanovsky, D. Cormier, H. J. de Vega, R. Holman et S. Prem
Kumar, Phys. Rev.  {\bf D57}, 2166, (1998),
(and references therein).

\bibitem{erickwu}  E. J. Weinberg and A. Wu, Phys. Rev. D36, 2474 (1987).

\bibitem{boyvega} D. Boyanovsky and H. J. de Vega, Phys. Rev. {\bf
D47}, 2343 (1993). 

\bibitem{rg} V. Branchina, P. Castorina and D. Zappal\`a,
Phys. Rev. {\bf D41}, 1948 (1990). 
A. Ringwald and C. Wetterich, Nucl. Phys. {\bf B334}, 506 (1990).
 
N. Tetradis and C. Wetterich,
Nucl. Phys. B 383, 197 (1992). N. Tetradis and
D. Litim, Nucl. Phys. B 464, 492 (1996)

\bibitem{polo} 
J. Alexandre, V. Branchina and  J. Polonyi, cond-mat/9803007. 
 J. Alexandre, V. Branchina and J. Polonyi, Phys. Rev. D58 (1998) 016002;
 hep-th/9709060; S.-B. Liao and J. Polonyi,  Phys. Rev. D51 (1995) 748.

\bibitem{tetradis} N. Tetradis and C. Wetterich, Int.J.Mod.Phys. A9
 (1994) 4029; Nucl.Phys. B422 (1994) 541; M. Reuter, N. Tetradis and
 C. Wetterich,  Nucl.Phys. B401 (1993) 567; 
 J. Berges, N. Tetradis, C. Wetterich, Phys.Lett. B393  (1997) 387.

\bibitem{root} R. G. Root, Phys. Rev.  D 10, 3322 (1974).
S. Coleman, R. Jackiw and H. D. Politzer,  Phys. Rev.  D 10, 2491 (1974).

\bibitem{aks} L. F. Abbott, J. S. Kang and H. J. Schnitzer,
Phys. Rev.  D 13, 2212 (1976).


\bibitem{barmo} W. A. Bardeen and M. Moshe, Phys. Rev.  D 28, 1372 (1983).


\bibitem{kibble} T. W. B. Kibble, J. Phys. A 9, 1387 (1976).

\bibitem{kibble2} M. B. Hindmarsh and T.W.B. Kibble,
  Rep. Prog. Phys. {\bf 58}:477 (1995). 

\bibitem{vilen} A. Vilenkin and E.P.S. Shellard, `Cosmic Strings and
  other Topological Defects',  Cambridge Monographs on Math. Phys. (Cambridge
  Univ. Press, 1994). 


\bibitem{zurek} W. H. Zurek, Nature 317, 505 (1985); Acta Physica 
Polonica B24, 1301 1993); Phys. Rep. 276, (1996). 


\bibitem{boylee}  D. Boyanovsky, D-S. Lee, and A. Singh, Phys. Rev.
{\bf D48}, 800 (1993).

\bibitem{rivers}  G.Karra and  R.J.Rivers, Phys.Lett. B414 (1997), 28;
 R.J.Rivers, 3rd. Colloque Cosmologie, Observatoire de Paris, June
 1995, p. 341 in the Proceedings edited by H J de Vega and
 N. S\'anchez, World Scientific.

A.J. Gill and R.J. Rivers, Phys. Rev. D51 (1995), 6949; 
G.J. Cheetham, E.J. Copeland, T.S. Evans, R.J. Rivers, 
 Phys. Rev.D47 (1993),5316. 

\bibitem{beilok} 
G. J. Stephens, E. A. Calzetta, B. L. Hu, S. A. Ramsey, Phys. Rev. D59
(1999) 045009. 
D. Ibaceta and E. Calzetta, hep-ph/9810301 (1998).

\bibitem{bray} A. J. Bray, Adv. Phys. {\bf 43}, 357 (1994). 

\bibitem{marco} C. Castellano and M. Zannetti, cond-mat/9807242;
C. Castellano, F. Corberi and M. Zannetti, Phys. Rev. E56, 4973
(1997); F. Corberi, A. Coniglio and M. Zannetti, Phys. Rev. E51, 5469 (1995). 

\bibitem{turok} N. Turok and D. N. Spergel, Phys. Rev. Lett. 66, 3093
(1991); D. N. Spergel, N. Turok, W. H. Press and B. S. Ryden,
Phys. Rev. D43, 1038 (1991).  

\bibitem{filipe} J. A. N. Filipe and A. J. Bray, Phys. Rev. E50, 2523
(1994); J. A. N. Filipe, (Ph. D. Thesis, 1994, unpublished).  


\bibitem{bowick} Relaxing the assumption of an instantaneous quench
and allowing for a time dependence of the cooling mechanism has been
recently studied by  M. Bowick and A Momen, Phys. Rev. D58 (1998) 085014.

 

\end{thebibliography}
\end{document}